\documentclass[12pt]{article}

\newlength{\abstractwidth}
\usepackage{amsfonts}
\usepackage{amssymb}
\usepackage{amsmath}
\usepackage{amsfonts}

\usepackage{epsf}
\usepackage{color}
\usepackage{graphicx}
\usepackage{tikz}
\usepackage{dsfont}

\usetikzlibrary{arrows,shapes,positioning}
\usetikzlibrary{decorations.markings}
\usepackage[rightcaption]{sidecap}
\tikzstyle arrowstyle=[scale=1]
\tikzstyle directed=[postaction={decorate,decoration={markings,
    mark=at position .65 with {\arrow[arrowstyle]{stealth}}}}]
\tikzstyle reverse directed=[postaction={decorate,decoration={markings,
    mark=at position .65 with {\arrowreversed[arrowstyle]{stealth};}}}]
\usetikzlibrary{positioning}

\usepackage{latexsym}
\usepackage{tikz}
\usetikzlibrary{fadings,decorations.pathmorphing}
\usetikzlibrary{patterns}
\tikzset{snake it/.style={decorate, decoration=snake}}
\usepackage{braket}
\usepackage{hyperref}
\usepackage{amsfonts}
\usepackage{amssymb}
\usepackage{amsthm}
\usepackage{pgfplots}

\renewcommand{\thefootnote}{\fnsymbol{footnote}}
\renewcommand{\thanks}[1]{\footnote{#1}}
\newcommand{\starttext}{
\setcounter{footnote}{0}
\renewcommand{\thefootnote}{\arabic{footnote}}}

\newcommand{\bea}{\begin{eqnarray}}
\newcommand{\eea}{\end{eqnarray}}
\newcommand{\be}{\begin{eqnarray}}
\newcommand{\ee}{\end{eqnarray}}

\DeclareMathOperator{\Vol}{Vol}

\def\t{\theta}
\def\l{\lambda}
\def\d{\delta}
\def\r{\rho}
\def\e{\epsilon}
\def\mc{\mathcal}
\def\a{\alpha}

\def\g{\gamma}
\date{}

\begin{document}
\starttext
\setcounter{footnote}{0}

	\begin{center}
		
		{\Large \bf  A note on  entanglement entropy and regularization in holographic interface theories }
		
		\vskip 0.4in
		
		{\large Michael Gutperle and  Andrea Trivella}
		
		\vskip .2in

	{ \sl 	Mani L. Bhaumik Institute for Theoretical Physics}\\
{\sl  Department of Physics and Astronomy}\\
{\sl University of California, Los Angeles, CA 90095, USA}

\vskip 0.05in

{\tt \small  gutperle@ucla.edu; andrea.trivella@physics.ucla.edu}

\vskip 0.5in
		
	\end{center}
		
		\bigskip
		
		\bigskip
		
\begin{abstract}
			\setlength{\baselineskip}{18pt}
We discuss  the computation of holographic entanglement entropy for interface conformal field theories. The fact that globally well defined Fefferman-Graham coordinates are difficult to construct makes the regularization  of the holographic theory challenging.  We introduce a simple new cut-off procedure, which we call  ``double cut-off" regularization.  We test the new cut-off procedure by comparing the results for holographic entanglement entropies  using other cut-off procedures and find agreement.  We also  study three dimensional  conformal field theories with a two dimensional interface. In that case the dual bulk geometry is constructed using warped geometry with an  $AdS_3$ factor.  We define  an effective central charge to the interface through the Brown-Henneaux formula for the $AdS_3$ factor. We investigate two concrete examples, showing that the same effective central charge appears in the computation of entanglement entropy and governs the conformal anomaly.

\end{abstract}
\newpage
\tableofcontents
\newpage

\baselineskip=16pt
\setcounter{equation}{0}
\setcounter{footnote}{0}

\section{Introduction}
\setcounter{equation}{0}
\label{intro}

The AdS/CFT correspondence provides the most well understood example of holography. The degrees of freedom of a theory of gravity in a geometry that includes an asymptotically $AdS$ space are encoded in the degrees of freedom of a dual conformal field theory, living on the boundary of the asymptotically $AdS$ space \cite{Maldacena:1997re,Gubser:1998bc,Witten:1998qj}.

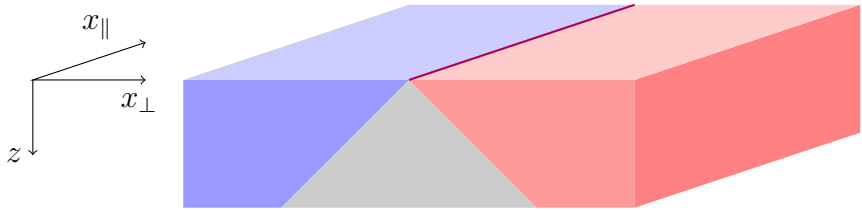
\begin{figure}[b]
	\centering
	\begin{tikzpicture}
	\fill[black!20!white]  (-1.7,-1.7)--(0,0)--(1.7,-1.7)--(-1.7,-1.7);
	\fill[blue!40!white] (-3,0)--(-3,-1.7)--(-1.7,-1.7)--(0,0);
	\fill[red!40!white] (3,0)--(3,-1.7)--(1.7,-1.7)--(0,0);
	\fill[red!50!white] (3,0)--(3,-1.7)--(6,-0.7)--(6,1);
	\fill[blue!20!white] (0,0)--(3,1)--(0,1)--(-3,0);
	\fill[red!20!white] (0,0)--(3,1)--(6,1)--(3,0);
	\draw[ thick, blue!30!red](0,0)--(3,1);
	\draw[<->] (0.5-4,0.5)--(-1-4,0)--(-3.5,0);
	\draw[->] (-5,0)--(-5,-1);
	\node[anchor=east] at (-5,-1) {$z$};
	\node[anchor=east] at (-3.2, -0.3) {$x_\perp$};
	\node[anchor=east] at (-3.8,0.7) {$x_\parallel$};
	\end{tikzpicture}
	\caption{The top surface represents the field theory side, the two different colors identify the two sides of the interface (purple line). The vertical dimension represents the holographic direction, there are two Fefferman-Graham coordinate patches (represented with different colors) that do not cover the entire bulk geometry. In the gray wedge originating from the interface the Fefferman-Graham coordinate expansion breaks down.} \label{figintro}
\end{figure}

The correspondence is mostly studied in the large $N$ and large t'Hooft coupling limit, when the the bulk side can be treated using semi-classical gravity.
For example, the Ryu-Takayanagi formula relates entanglement entropy on the field theory side to the area  of the minimal bulk co-dimension two surface anchored at the boundary of $AdS$ on the entangling surface \cite{Ryu:2006bv}
\begin{equation}\label{RTformula}
S_{EE}=\frac{A_{\text{min}}}{4 G_N}.
\end{equation}

One should note that the entanglement entropy on the field theory and gravity side are infinite and both require regularization. On the CFT side the divergence comes from the short distance degrees of freedom entangled across the entangling surface, for this reason a UV cut-off is required. On the gravity side the divergence arises from the fact that the minimal surface is anchored on the boundary of the asymptotic $AdS$ space, which has an infinite volume. For that reason we need to regulate it by introducing a cut-off on the holographic coordinate, this process is called holographic renormalization (for a review see \cite{Skenderis:2002wp}). The regularization is based on the fact that an asymptotically $AdS$ metric can be expressed in terms Fefferman-Graham coordinates \cite{FG}.
\begin{eqnarray}\label{FGcoord}
ds^2&=&\frac{dz^2}{z^2}+\frac{1}{z^2}g_{i j}(x, z) dx^i dx^j
\end{eqnarray}
Where $g_{i j}(x, z)$ has a leading $z$ independent term and terms falling off as $z\to 0$, whose exact form depend on the dimensionality and details of the theory. 

The boundary of the asymptotic $AdS$ metric is located at $z=0$ and the  theory is regulated by 
imposing a cut-off at $z=\d$.

Unfortunately the construction of  Fefferman-Graham coordinates which cover all of the boundary can be difficult. One  example are systems with an interface (ICFT) or a defect (DCFT).
In the present paper we consider holographic interface or defect solutions which are commonly known as Janus solutions, where one solved the bulk gravitational equations for a metric which is warped with an $AdS$ factor. For some other approaches to describe interface, defect or boundary CFTs holographically see e.g. \cite{Takayanagi:2011zk,Fujita:2011fp,Aharony:2003qf,Karch:2000gx}.

In these cases  the small $z$ expansion used for the Fefferman-Graham construction turns out to be an expansion in small $z/x_{\perp}$, where $x_{\perp}$ denotes the field theory direction perpendicular to the defect. This dependence is dictated by scale invariance. The expansion breaks down close to the defect, where $x_\perp \rightarrow 0$. Thus there is a wedge bulk region originating from the defect that cannot be covered. In the case of a co-dimension one defect we have two different Fefferman-Graham coordinates patches that cover some portion of the bulk on the two sides of the defect and a region just behind the defect that cannot be covered. A~schematic representation is  given in figure \ref{figintro}.

This problem has been faced in literature in different ways. The authors of \cite{Estes:2014hka} connected the two Fefferman-Graham patches with an arbitrary curve, showing that any universal quantity would not depend on the details of this curve. To avoid dealing with Fefferman-Graham coordinates the authors of \cite{Bak:2016rpn} simply imposed a cut off on the factor of the metric that diverges as one moves to the boundary. We refer to this regularization procedure as ``single cut-off regularization''.

Recently,  a  third regularization procedure has been used in literature in the computation of the quantum information metric of a conformal theory which is deformed by a primary operator.  Such a set up shares a lot of similarities with a DCFT \cite{MIyaji:2015mia,Bak:2015jxd,Trivella:2016brw} since it is natural to express the bulk metric using an $AdS$ slicing. In such coordinates one encounters a divergence associated to the infinite volume of the $AdS$ slice and a divergence associated to the coordinate that slices the bulk geometry. It is then natural to  introduce two cut offs. We name this regularization procedure ``double cut-off regularization''. Note that an analogous cutoff was also used to regulate  holographic duals of surface operators, i.e. defects of higher co-dimensionality in 	\cite{Gentle:2015jma,Gentle:2015ruo}.

The purpose of this paper is to study the double cut-off regularization in more detail. We will test it against several examples to show that it provides the same results as the other regularization methods  but involve  much simpler computations. 

The paper is organized as follows: after reviewing and discussing the main features of different cut-off procedures in section \ref{reg}, we move on to discuss specific examples provided by ICFTs with a co-dimension one planar interface. In section \ref{higherdim} we discuss systems 
with an interface extended along at least two spatial dimensions. The computation of the 
entanglement entropy in these cases has been carried out in \cite{Estes:2014hka} and we find 
agreement between the calculations which utilize the old and new regularization methods. In 
section \ref{3dim} we focus on three dimensional CFTs with a two  dimensional conformal 
interface. The bulk geometry dual to this systems is given by a warped space with a $AdS_3$ 
factor. We  associate to the interface an effective central charge through the 
Brown-Henneaux formula for the $AdS_3$ factor. We study two concrete examples, showing that the effective central charge obtained holographically appears also in the computation of the entanglement entropy and it is the same quantity that governs the conformal anomaly associated with a two dimensional CFT living on the interface.

\section{Regularization prescriptions}\label{reg}
\setcounter{equation}{0}

In this paper we mainly focus on the computation of entanglement entropy for a ball shaped region in a CFT with a co-dimension one interface. This quantity is divergent because of the UV degrees of freedom entangled across the entangling surface. The regularization is achieved by introducing a UV cut-off. Once this is done if we want to isolate the interface contribution  we need to subtract the entanglement entropy for the vacuum of the theory without interface. In this way we are able to compute a quantity that is intrinsic to the interface. To better explain this statement let us discuss in detail the divergence structure of entanglement entropy. For the vacuum state of a pure CFT and a ball shaped region of radius $R$ we have:
\begin{equation}
S_{EE}=A_{d-2} \frac{R^{d-2}}{\d^{d-2}}+...+\begin{cases} A_{1} \frac{R}{\d}+s_0 &\text{  if $d$ is odd}\\
A_{2} \frac{R^2}{\d^2}+ s \log(2R/\d)+\tilde s_0& \text{  if $d$ is even}
\end{cases}
\end{equation}
where we have introduced the UV cut-off $\d$ \cite{Srednicki:1993im}. Notice that in odd dimensions a rescaling of the cut-off does not affect constant $s_0$, while in even dimension it is the coefficient of the logarithmic term, $s$, that is not sensitive to any rescaling of $\d$. For this reason $s$ and $s_0$  are independent of regularization and   are  universal. Let us discuss how the presence of a defect affects the structure of entanglement entropy. For definiteness we  start with the vacuum state of an even dimensional CFT. We then turn on a co-dimension one interface that breaks the full conformal symmetry group $SO(2,d)$ down to $SO(2,d-1)$, interpreted as the conformal symmetry restricted to the interface. When this is done we expect the entanglement entropy to show terms typical of both even and odd dimensional CFTs \cite{Jensen:2013lxa}. That creates a problem in isolating the universal term characterizing the interface. In fact since the interface is odd dimensional we expect that the universal term should be a constant, however since the original CFT is even dimensional we have a logarithmic term in the divergence structure of the entanglement entropy and we are free to change the additive constant by a rescaling of the cut-off $\d$. The way to bypass this problem is to use the same cut-off for both the pure CFT and the ICFT, once that is done we can isolate the interface contribution by subtracting the vacuum component. We refer to this procedure as vacuum subtraction. 

\begin{figure}[t]
	\centering
	\begin{tikzpicture}
	\draw[color=blue] (-3,0)--(0,0);
	\draw[color=red] (0,0)--(3,0);
	\draw[color=blue!20!red] (0,0) ..controls (0.5,-0.25) and (1,-0.5).. (2.8,-0.6);
	\draw[color=blue!30!red] (0,0) ..controls (0.5,-0.5) and (1,-1).. (2.2,-1.2);
	\draw (0,0)[color=blue!40!red] ..controls (0.2,-0.7) and (1,-1.6).. (1.3,-1.8);
	\draw[color=blue!50!red] (0,0) --(0,-2);
	\draw[color=blue!60!red] (0,0) ..controls (-0.2,-0.7) and (-1,-1.6).. (-1.3,-1.8);
	\draw[color=blue!70!red] (0,0) ..controls (-0.5,-0.5) and (-1,-1).. (-2.2,-1.2);
	\draw[color=blue!80!red] (0,0) ..controls (-0.5,-0.25) and (-1,-0.5).. (-2.8,-0.6);
	\node[anchor=south] at (0,0) {$Z=0$};
	\node[anchor=south] at (2.2,0) {$x=\infty$};
	\node[anchor=south] at (-2.2,0) {$x=-\infty$};
	\node[anchor=west] at (2,-1.2) {\textcolor{blue!30!red}{$AdS_{d}$}};
	\node[anchor=east] at (-2,-1.2) {\textcolor{blue!70!red}{$AdS_{d}$}};
	\end{tikzpicture}
	\caption{Schematic representation of the $AdS_d$ slicing of the bulk geometry $\mc M$. Each colored line corresponds to a single $AdS_d$ slice located at a fixed value of the coordinate $x$.} \label{figslicing}
\end{figure}
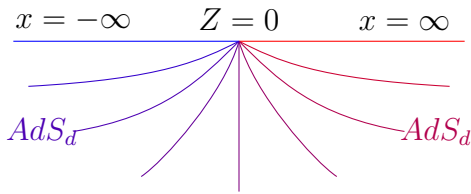

Now that we have discussed regularization and vacuum subtraction on the CFT side of the duality let's focus on the bulk side, where all the computations will be performed. First of all we need to identify a bulk geometry dual to the interface CFT. This is realized by a metric that is invariant under $SO(2,d-1)$ transformations. The natural way to do that is to consider a bulk geometry $\mc M$ that can be written in $AdS_{d}$ slices:
\begin{equation}\label{metric}
ds^2=A(x,y^a)^2 g_{AdS_{d}}+\r (x, y^a)^2 dx^2+G_{bc}(x,y^a) dy^b dy^c.
\end{equation}
The coordinate $x$ is taken to be non compact and as $x\rightarrow \pm \infty$ we have $A(x,y^a)\approx L_{\pm} \exp(\pm  x+c_{\pm})/2$ and $\r(x,y^a)\approx 1$ such that the $AdS_{d}$ gets enhanced to $AdS_{d+1}$. Unless otherwise stated we will work in Poincar\'e coordinates for the $AdS_{d}$ slices 
\begin{equation}\label{poincare}
 g_{AdS_{d}}=\frac{1}{Z^2}(Z^2-dt^2+dr^2+r^2g_{\mathbb{S}^{d-3}}).
\end{equation}
The boundary is approached in different ways. Taking $x \rightarrow  \pm \infty$ we recover the CFT region on the right/left side of the interface, while taking $Z \rightarrow 0$ we approach the CFT on the interface itself. A schematic illustration is given in figure \ref{figslicing}.
 
We will now describe  how to regularize divergent quantities on the bulk side using  three different methods.
\begin{figure}[b]
	\centering
	\begin{tikzpicture}
	\fill[black!20!white]  (-1.7,-1.7)--(0,0)--(1.7,-1.7)--(-1.7,-1.7);
	\fill[blue!40!white] (-7,0)--(-7,-1.7)--(-1.7,-1.7)--(0,0);
	\fill[red!40!white] (7,0)--(7,-1.7)--(1.7,-1.7)--(0,0);
	\fill[blue!30!red] (0,0) circle (0.15cm);
	\draw[thick,blue] (-7,-0.5)--(-0.5,-0.5);
	\draw[thick, red] (7,-0.5)--(0.5,-0.5);
	\draw[thick] (-0.5,-0.5) arc (225:315:0.707);
	\node[above] at (0,0) {\textcolor{blue!30!red}{interface}};
	\node[above] at (-6,-1.2) {\textcolor{blue}{Left}};
	\node[below] at (-6,-1) {\textcolor{blue}{FG patch}};
	\node[above] at (6,-1.2) {\textcolor{red!60!black}{Right}};
	\node[below] at (6,-1) {\textcolor{red!60!black}{FG patch}};
	\node[above] at (-3,-0.5) {\textcolor{blue}{$z=\d$}};
	\node[above] at (3,-0.5) {\textcolor{red!60!black}{$z=\d$}};
	\end{tikzpicture}
	\caption{Schematic representation of the Fefferman-Graham regularization. Where the Fefferman-Graham coordinates are available (red and blue regions) the cut off surface is chosen to be $z=\e$. In the middle region a Fefferman-Graham coordinate patch is not available. The cut off surface for this region is an arbitrary curve that continuously interpolates between the left and right patches, this is represented by a black arc in the picture.} \label{FGregpic}
\end{figure}
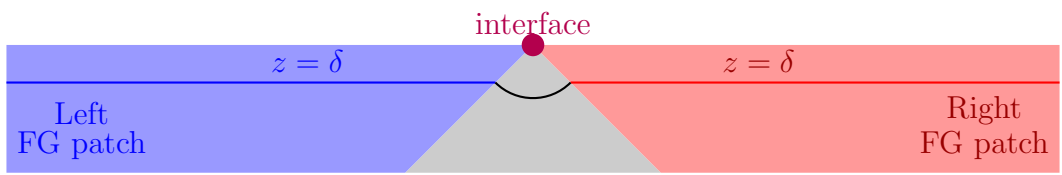
\begin{itemize}
\item \textbf{Fefferman-Graham regularization}: The traditional approach is to make use of Fefferman-Graham coordinates. As mentioned in the introduction this is problematic in a bulk geometry that is dual to a CFT with a defect or interface. There are two Fefferman-Graham patches which do not overlap, so one cannot simply glue them together. A possibility is then to interpolate with an arbitrary curve between these two patches, this is the approach used in \cite{Estes:2014hka} where the authors were able to compute universal quantities that do not depend on the interpolating curve. Even though this approach is very rigorous it requires a heavy computational effort. For this reason we want to explore other regularization procedures. A schematic representation of this procedure is given in figure \ref{FGregpic}.

\item \textbf{ single cut-off regularization}: we follow the idea of \cite{Bak:2016rpn}, regularizing all the divergent integrals by putting a cut-off at $Z/A(x)=\d/L_{\pm}$. This is motivated by the study of pure $AdS_{d+1}$. In fact for pure $AdS_{d+1}$ with unit radius one has $A(x)=\cosh x$, we can then change coordinates to recover Poincar\'e $AdS_{d+1}$ by choosing:
\begin{eqnarray}
z=\frac{Z}{\cosh x} &  & \tilde x= Z \tanh x,
\end{eqnarray}
where $z$ is the holographic coordinate and $\tilde x$ is the coordinate perpendicular to the fictitious interface. The natural cut-off procedure $z=\d$ corresponds, in the $AdS_{d}$ slicing coordinates, to $ Z/A(x)=\d$. For the interface solution which can be viewed as a deformation away from the $AdS$ vacuum  we keep the same regularization procedure.

\item \textbf{double cut-off regularization}: this procedure is based on the observation that, after one performs the vacuum subtraction, one should be left with a quantity that is intrinsic to the interface. In that sense a cut-off should be imposed not on the full bulk geometry but on the $AdS_{d}$ slices, at $Z=\d$. Of course that cut-off does not regulate all the possible divergences, since the metric factor in (\ref{metric}) diverges as $A(x)\approx L_{\pm}\exp(\pm x + c_{\pm})/2$ as $x \rightarrow \pm \infty$. What one should do is to introduce a second cut-off $\e$, such that $ A(x)=L_{\pm}\e^{-1} $, that regulates any $x$ dependent divergence.  Once we subtract the vacuum contribution to the particular physical quantity in consideration we will be allowed to take $\e\rightarrow 0$, the result will be $\e$ independent. To sum up, the double cut-off procedure makes use of two cut-offs $\d$ and $\e$. $\d$ is interpreted as a physical cut-off in the usual sense, it regulates the bulk divergence associated to the $AdS_{d}$ integration and it is interpreted as a UV cut-off for the degrees of freedom localized on the interface. On the other side the $\e$ cut-off is  a  purely mathematical tool. It is used only to make any quantity that appears in the intermediate steps finite, any physical quantity should be $\e$ independent. 
\end{itemize}

\begin{figure}
	\centering
	\begin{tikzpicture}
	\fill[fill=blue!20!white] (-1.5,0)--(1.5,0)--(4,1.5)--(1,1.5)--cycle;
	\fill[fill=red!20!white] (4.5,0)--(1.5,0)--(4,1.5)--(7,1.5)--cycle;
	\draw[ thick, blue!30!red] (1.5,0)--(4,1.5);
	\node at (-1.65,-0.5) {
		\begin{axis}[axis x line=none, axis y line=none, axis z line=none,samples=20,domain=0:360,y domain=0:90,
		xmin=-1.2,xmax=1.2,ymin=-1.2,ymax=1.2,zmin=0,zmax=1.2]
		\addplot3[surf, color=green!70!blue, opacity=0.1,fill opacity=0.5, faceted color=black 
		]
		({0.7*cos(x)*cos(y)}, {0.7*sin(x)*cos(y)}, {0.7*sin(y)});
		\end{axis}};
		\draw[<->] (6.25,0.75)--(-1+6,0)--(-3.5+10,0);
		\draw[->] (5,0)--(5,1);
		\node[anchor=east] at (5,1) {$z$};
		\node[anchor=east] at (6.7, -0.3) {$x_\perp$};
		\node[anchor=east] at (7,0.7) {$x_\parallel$};
	\end{tikzpicture}
	\caption{Representation of a time slice of the field theory side. Two regions (blue and red) are separated by a interface (purple). We compute the holographic entanglement entropy for a ball centered on the interface. The Ryu-Takayanagi surface is represented in green.} \label{setup}
\end{figure}
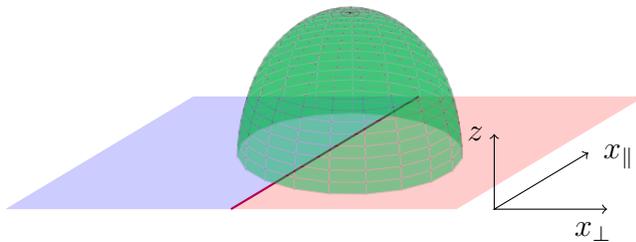

This discussion applies to any divergent quantities that can be computed  in a holographic  ICFT. Let us now focus on the computation of holographic entanglement entropy. We take the entangling surface to be a ball shaped region of radius $R$ centered on the interface (see figure \ref{setup}). The holographic entanglement entropy for these systems has been studied in \cite{Jensen:2013lxa}, where the authors were able to show that the RT surface is simply given by $r^2+Z^2=R^2$, giving the following expression for the entanglement entropy
\begin{equation}\label{HEE}
S=\frac{\Vol(\mathbb{S}^{d-3})R}{4 G_N} \int dy^a dx dZ \sqrt{\text{det}G} \r A^{d-2}\frac{(R^2-Z^2)^{(d-4)/2}}{Z^{d-2}}.
\end{equation}
This equation can be adapted also for $d=3$ by taking $\Vol(\mathbb{S}^0)=2$.

Let us discuss how to regulate the entanglement entropy using the single and double cut-off regularizations. For the double cut-off procedure we cut-off the $x$ integral at $x= x'_{\pm}$, defined as the two roots of $A(x')=L_{\pm}\e^{-1}$. In most examples $A(x)^2$ is an even function, in that case $x'_+=-x'_-$, we can then focus only on $x \in [0, x'_+]$ and we will drop the subscript. Generally speaking the form of $A$ might be very complicated, however since $\e$ eventually goes to zero we can assume $x'$ large, allowing us to find $x_{\pm}'=\pm \left(\log (2 \e)-c_{\pm}\right)$. We introduce a cut-off for the $Z$  integration at $Z=\d$. We then get:
\begin{equation}
 \Delta S=\frac{\Vol(\mathbb{S}^{d-3})R}{4 G_N} \left( \int_{\d}^{R}dZ\frac{(R^2-Z^2)^{(d-4)/2}}{Z^{d-2}}\right) \int dy^a \Delta \left(\int_{x_-}^{x_+}dx \sqrt{\text{det}G} \r A^{d-2}\right),
\end{equation}
where the $\Delta$ symbol denotes the vacuum subtraction. At this point we will take $\d,\e \rightarrow 0$. The divergence will come exclusively from the $Z$ integral and the result will be $\e$ independent.

We will now discuss the single cut-off procedure for the entanglement entropy. In this case we put a cut-off at $Z/A(x)=\d/L_{\pm}$. We will always proceed by performing the $x$ integral first and then the $Z$ integral. To do so we start by fixing $Z$ and integrating in $x$ over $[\tilde x_-, \tilde x_+]$, where $\tilde x_{\pm}$ are the solutions to $Z/A(x)=\d/L_{\pm}$. At this point we might be tempted to take $\d$ small, however that is not possible. The reason for it is that the integration over $Z$ runs over $[\text{min}(A) \d/L_{\pm}, R]$, where $\text{min}(A)$ denotes the minimum of $A$ (in most examples that corresponds to $x=0$). Nonetheless we can expand $\exp(\tilde x_{\pm})$ as a Laurent series in $\d/Z$. Once this is done we will proceed to the integration, whose details depend on the concrete examples we will examine.

Notice that one could work in different coordinates than (\ref{metric}). In particular one could change coordinates from  $x$ to another coordinate, say $q$. The function $A(x)$ will then be replaced with another function, say $B(q)$. In that case the regularization procedures just described will go through without any change, one would simply put a cut-off for the $q$ integration at $B(q)=L_{\pm} \e^{-1}$ for the double cut-off procedure and at $B(q)=L_{\pm}Z \d^{-1}$ for the single cut-off procedure.

\section{Higher Dimensional Examples}\label{higherdim}
\setcounter{equation}{0}

In this section we discuss the computation of the holographic entanglement entropy for ICFT that present an interface extended on at least two  spatial dimension. We will leave the discussion of lower dimensional cases in section \ref{3dim}. 
\subsection{Supersymmetric Janus}
In this section we discuss the entanglement entropy for a ball shaped region for a Yang-Mills interface that preserves 16 supercharges \cite{D'Hoker:2007xz,D'Hoker:2006uv}. That is realized in the bulk by a metric that explicitly exhibits $SO(2,3)\times SO(3)\times SO(3)$ symmetry where the first factor is associated to the conformal symmetry preserved on the interface and the other two factors are related to unbroken R-symmetry. The full supergravity solution  also has the dilaton, the three-form and the five-form are turned on in the bulk, see \cite{D'Hoker:2007xz} for details.  In the following we will only need the metric which is given by:
\begin{equation}
ds^2= f_4^2 ds^2_{AdS_4}+\r^2 dv d \bar{v}+f_1^2 ds_{\mathbb{S}^2}+f_2^2 ds_{ \mathbb{\tilde S}^2}.
\end{equation}
The coordinates $v$ and $\bar v$ parametrize a two dimensional Riemann surface with boundary. The functions $f_4$, $f_2$, $f_1$ and $\r$ depend on $v$, $\bar v$ and they can be obtained from two functions $h_1$ and $h_2$ in the following way:
\begin{eqnarray}
f_4^8&=&16 \frac{F_1 F_2}{W^2}, \quad \quad  \quad \; \r^8\; = \; \frac{2^8 F_1 F_2 W^2}{h_1^4 h_2^4} \nonumber\\
f_1^8&=& 16 h_1^8 \frac{F_2 W^2}{F_1^3}, \quad \quad  f_2^8\; = \;16 h_2^8 \frac{F_1 W^2}{F_2^3}
\end{eqnarray}
where
\begin{eqnarray}
F_i&=& 2 h_1 h_2 |\partial_v h_i|^2-h_i^2 W, \quad \quad   W=\partial_v \partial_{\bar v} (h_1 h_2). 
\end{eqnarray}
For the supersymmetric Janus solution we have:
\begin{eqnarray}
h_1&=&- i \alpha_1 \sinh\left(v-\frac{\Delta \phi}{2}\right)+c.c.\nonumber \\
h_2&=&\alpha_2 \cosh\left(v+\frac{\Delta \phi}{2}\right)+c.c.
\end{eqnarray}
with $v=x+i y$ and $x\in\mathbb{R}$ and $0\leq y \leq \pi/2$. The asymptotic regions located at $x \rightarrow \pm \infty$ correspond to the two sides of the interface, where the dilaton assumes different values corresponding to different values of the Yang Mills coupling constant $g_{YM}^{\pm}$. The constants $\a_1,\a_2$ and $\Delta \phi$ are reals and they are related to the $AdS$ radius and to the Yang Mills coupling constant by:
\begin{eqnarray}
L^4&=&16 |\alpha_1 \alpha_2| \cosh \Delta\phi \nonumber\\
(g_{YM}^{\pm})^2&=&4 \pi \left|\frac{\a_2}{\a_1}\right| e^{\pm \Delta \phi}
\end{eqnarray} 

Equation (\ref{HEE}) gives the following expression for the entanglement entropy of a ball shaped region centered on the interface:
\begin{equation}
S=\frac{\Vol(\mathbb{S}^1)\Vol(\mathbb{S}^2)^2 R L^8}{4 G_N} \int_{0}^{\pi/2} dy \sin^2 y \cos^2 y \int_{\text{cut-off}}^{R}\frac{d Z}{2 Z^2} \int_{0}^{\text{cut-off}}2\left(1+\frac{\cosh 2x}{\cosh \Delta \phi}\right) dx.
\end{equation}

We now need to specify the cut-off procedure. We dedicate the next two sections to two different regularizations.
\subsubsection*{Single cut-off}
For the single cut-off procedure we have:
\begin{equation}
\frac{f_4^2}{Z^2}=\frac{L^2}{\delta^2}. 
\end{equation}
We start by fixing $Z$ letting $x$ varying from $0$ to $\tilde x$, with $\tilde x$ defined by:
\begin{equation}
f_4(\tilde x)=\frac{L Z}{\d}.
\end{equation}
Notice that even though we are going to let $\d\rightarrow 0$, we cannot assume $\tilde x$ to be large, since $z \in [\delta f_4(x=0)/L, R]$. Nonetheless we can expand $\tilde x=f^{-1}\left( \frac{LZ}{\d}\right)$ in Laurent series of $\frac{\d}{Z}$. We have:
\begin{equation}
e^{2 \tilde x}=2^{3/2} \left( \cosh \Delta \phi \right) \left(\frac{Z}{\d}\right)^2\left(1+\sum_{k=2}^{\infty} c_k (y) \left(\frac{\d}{Z}\right)^k\right),
\end{equation}
thus:
\begin{equation}
\tilde x= \frac{1}{2}\log\left(2^{3/2} \cosh \Delta \phi\left(\frac{Z}{\d}\right)^2\right)+\sum_{k=2}^{\infty} c_k (y) \left(\frac{\d}{Z}\right)^k.
\end{equation}
Of course the coefficients in the sum are going to be different with respect to the one of the previous equations, but since we are not really interested in those coefficients   we will adopt a loose notation.
We can now perform the integral over $x$:
\begin{eqnarray}
\mathcal{P}_2(y,Z)&\equiv& \int_{0}^{{\tilde x(Z,y)}}2\left(1+\frac{\cosh 2x}{\cosh \Delta \phi}\right) dx\nonumber \\
&=& \log\left(\frac{2^{3/2} \cosh \Delta \phi Z^2}{\d^2}\right)+\frac{2^{3/2}Z^2}{\d^2}+ \sum_{k=2}^{\infty} c_k (y) \left(\frac{\d}{Z}\right)^k.
\end{eqnarray}
We proceed with the integration over $Z$:
\begin{eqnarray}
\int_{f_4(x=0)\d/L }^{R}\frac{R dZ}{2 Z^2} \mathcal{P}_2(y,Z)&=& \frac{\sqrt{2} R^2}{\d^2}-1-\log\left(\frac{R \sqrt{\cosh\Delta\phi} 2^{3/4}}{\d}\right)\nonumber\\
&&+\frac{R c_{-1}(y)}{\d}+ \sum_{k=2}^{\infty} c_k (y) \left(\frac{\d}{R}\right)^k
\end{eqnarray}
Integrating over $y$ and taking $\d \rightarrow 0$ leads to:
\begin{equation}
S(\Delta\phi)= \frac{\pi \Vol(\mathbb{S}^1)\Vol(\mathbb{S}^2)^2 L^8}{64 G_N} \left( \frac{\sqrt{2} R^2}{\d^2}-1-\log\left(\frac{R \sqrt{\cosh\Delta\phi} 2^{3/4}}{\d}\right)+\frac{R C}{\d}\right),
\end{equation}
for some constant $C$. Subtracting the vacuum contribution leads to\footnote{Note that we need to keep the $AdS$ radius fixed when we perform the vacuum subtraction.}:
\begin{equation}\label{delsa}
\Delta S=\frac{\pi \Vol(\mathbb{S}^1)\Vol(\mathbb{S}^2)^2 L^8}{64 G_N} \left(-\frac{1}{2}\log \cosh \Delta \phi+\frac{D R}{\d}\right),
\end{equation}
for some constant $D$, however  note  that $D$ is non universal. The universal contribution  is given  by the first term in (\ref{delsa}):
\begin{equation}
\Delta S_{\text{UNIV}}=-\frac{\pi \Vol(\mathbb{S}^1)\Vol(\mathbb{S}^2)^2 L^8}{128 G_N}\log \cosh \Delta \phi.
\end{equation}
\subsubsection*{Double cut-off}
We introduce  two different cut-offs $\delta$ and $\e$. We will use $\d$ to regulate the integration over $Z$ and $\e$ to regulate the integration over $x$. Remember that by vacuum subtraction we are going to obtain a result that is  $\e$-independent.

Let's start with the $x$ integration. We regularize it by cutting off the integral at $x= x'$, where $x'$ is defined by:
\begin{equation}
\frac{L^2}{f_4^2( x')}=\frac{1}{\e^2}.
\end{equation}
Notice that since $\e\rightarrow 0$, $x' \rightarrow \infty$, thus we can use the following asymptotic expression for $f_4$:
\begin{equation}
f_4^8(x)\approx 4 \left(\frac{\alpha_1 \alpha_2}{\cosh \Delta\phi}\right)^2 e^{8 x}.
\end{equation} 
We get:
\begin{equation}
x'=\frac{1}{2}\log\left(\frac{2^{3/2} \cosh \Delta\phi}{\e^2}\right).
\end{equation}
We then have:
\begin{equation}
\int_{0}^{{x'}}2\left(1+\frac{\cosh 2x}{\cosh \Delta\phi}\right) dx= \log\left(\frac{2^{3/2} \cosh \Delta\phi}{\e^2}\right)+\frac{2^{3/2}}{\e^2}+\mathcal{O}(\e^2).
\end{equation}
For the $Z$ integration we put a cut-off at $Z=\d$. We have:
\begin{equation}
R \int_{\d}^{R}\frac{d Z}{2 Z^2}=\frac{R}{2 \d}-\frac{1}{2}.
\end{equation}
The $y$ integration is finite and gives a $\pi/16$ factor. We obtain:
\begin{equation}
S(\Delta\phi)=\frac{\pi \Vol(\mathbb{S}^1)\Vol(\mathbb{S}^2)^2 L^8}{64 G_N}\left(\frac{R}{2 \d}-\frac{1}{2}\right)\left(\log\left(\frac{2^{3/2} \cosh \Delta\phi}{\e^2}\right)+\frac{2^{3/2}}{\e^2}+\mathcal{O}(\e^2)\right).
\end{equation}
Remember that in ICFT the physical information can be extracted only after a background subtraction. We obtain:
\begin{equation}
\Delta S=\frac{\pi \Vol(\mathbb{S}^1)\Vol(\mathbb{S}^2)^2 L^8}{64 G_N} \left(\frac{R}{2\d}-\frac{1}{2}\right)\log \cosh \Delta\phi.
\end{equation}
The universal contribution is
\begin{equation}
\Delta S_{\text{UNIV}}=-\frac{\pi \Vol(\mathbb{S}^1)\Vol(\mathbb{S}^2)^2 L^8}{128 G_N}\log \cosh \Delta\phi.
\end{equation}\\

Notice that we get the same result independently of the regularization procedure adopted. Moreover our result matches the expression found in literature using the Fefferman-Graham regularization \cite{Estes:2014hka}.  

\subsection{Non Supersymmetric Janus}\label{nonsusy}
The Non Supersymmetric Janus \cite{Bak:2003jk,D'Hoker:2007xy} is a solution of type IIB supergravity where the vacuum solution $AdS_5\times \mathbb{S}^5$ is deformed into the following metric
\begin{equation}
ds^2=L^2(\g^{-1} h(\xi)^2 d\xi^2+h(\xi) ds^2_{AdS_4})+L^2 ds_{\mathbb{S}^5}^2,
\end{equation}
where 
\begin{equation}
h(\xi)=\g \left(1+\frac{4 \g-3}{\wp(\xi)+1-2\g}\right)
\end{equation}
and $\wp$ is the $\wp$-Weierstrass function obeying $(\partial\wp)^2=4 \wp^3-g_2\wp-g_3$, with $g_2=16\g(1-\g)$ and $g_3=4(\g-1)$. The deformation depends on a real number $\g \in [3/4,1]$ called Janus deformation parameter. $\g=1$ corresponds to the vacuum solution. The metric is supported by a non trivial dilaton and RR five-form. This solution breaks all supersymmetries. Notice that $h(\xi)$ diverges as $\xi \rightarrow \pm \xi_0$, defined by $\wp(\xi_0)=2 \g-1$. The dilaton takes two different values in these asymptotic regions and the metric asymptotes to $AdS_5\times \mathbb{S}^5$. We interpret the bulk configuration as being dual to a deformation of $\mc N=4$ SYM, where an interface is present and the Yang Mills coupling constant takes different values on the two sides of the interface.  

Once the metric is available we can use equation (\ref{HEE}) to write the entanglement entropy for a ball shaped region of radius $R$ centered on the interface. We have:
\begin{equation}
S=\frac{\Vol(\mathbb{S}^1)\Vol(\mathbb{S}^5)R L^8}{4 G_N}\int_{\text{cut-off}}^{R}\frac{dZ}{Z^2}\int_{0}^{\text{cut-off}}d\xi\frac{2 h(\xi)^2}{\sqrt{\g}}.
\end{equation}
We now discuss in detail the two regularization procedures explained in \ref{reg}.
\subsubsection*{Single cut-off}
We introduce the cut-off $\d$ by
\begin{equation}
\frac{h(\xi)}{Z^2}=\frac{1}{\d^2}.
\end{equation}
We start with the integration over $\xi$. The cut-off for the $\xi$ integral is given by $\tilde \xi=h^{-1}(\frac{Z^2}{\d^2})$. Notice that we cannot simply take $\d$ small, since eventually $\d/Z$ is going to be $\mathcal{O}(1)$ when performing the $Z$ integral. Nonetheless we can perform a Taylor expansion in $\d/Z$, we find:
\begin{equation}
\tilde \xi=\xi_0-\frac{\sqrt{\g}\d^2}{2 Z^2}+\sum_{k=4}^{\infty} c_k \left(\frac{\d}{Z}\right)^{k}.
\end{equation} 
We then get:
\begin{equation}
\mathcal{P}_1(Z)\equiv\int_{0}^{\tilde \xi}d\xi\frac{2 h(\xi)^2}{\sqrt{\g}}=\frac{Z^2}{\d^2}+\log\left(\frac{2Z}{\d}\right)+\mathcal{B}+\sum_{k=1} \tilde c_k \left(\frac{\d}{Z}\right)^{k},
\end{equation}
for some coefficient $\tilde c_k$ and
\begin{equation}\label{Ci}
\mathcal{B}=-\frac{1}{4}-(\zeta(\xi_0)-\sqrt{\g})\xi_0+\frac{1}{2} \log\left(\frac{\sigma(2 \xi_0)}{2 \sqrt{\g}}\right)-\frac{\sqrt{\g}}{2}\zeta(2 \xi_0).
\end{equation}

We have now to perform the $Z$ integral, in particular $Z \in [\d \sqrt{h(0)},R]$:
\begin{equation}
\int_{\d \sqrt{h(0)}}^{R} \frac{dZ R}{Z^2} \mathcal{P}_1(Z).
\end{equation}
Let's look at the last term of $\mathcal{P}$. When we integrate the generic $k$-th term we obtain two terms, one behaving like $\d^k$ and the other as $\d^{-1}$, this means that the third term in $\mathcal{P}$ contribute to the divergence structure of $S$ with a term of the form $c/\d$. Let's now focus on the remaining terms, the integration is straightforward, one gets
\begin{equation}
S(\g)=\frac{\Vol(\mathbb{S}^1)\Vol(\mathbb{S}^5) L^8}{4 G_N}\left(\frac{R^2}{\d^2}+\frac{RC_\g}{\d}+\log\left(\frac{\d}{2 R}\right)-1-\mathcal{B}\right),
\end{equation} 
where we have dropped the terms that vanish as we take $\d\rightarrow 0$.\\
The vacuum entanglement entropy is given by taking $\g=1$:
\begin{equation}
S(\g=1)=\frac{\Vol(\mathbb{S}^1)\Vol(\mathbb{S}^5) L^8}{4 G_N}\left(\frac{R^2}{\d^2}+\frac{RC_1}{\d}+\log\left(\frac{\d}{2 R}\right)-1+\frac{1}{2}\right).
\end{equation} 
We then have:
\begin{equation}
\Delta S=-\frac{\Vol(\mathbb{S}^1)\Vol(\mathbb{S}^5) L^8}{4 G_N}\left(\frac{R(C_1-C_\g)}{\d}+\mathcal{B}+\frac{1}{2}\right),
\end{equation}
the universal contribution is given by:
\begin{equation}\label{EEnonsusy}
\Delta S_{\text{UNIV}}=-\frac{\Vol(\mathbb{S}^1)\Vol(\mathbb{S}^5) L^8}{4 G_N}\left(\mathcal{B}+\frac{1}{2}\right).
\end{equation}
\subsubsection*{Double cut-off}
We regulate the $Z$ integral and the $\xi$ integral using two different cut-offs. Let's start with the integral over $\xi$. This integral is divergent because $h(\xi)$ blows up at $\xi=\xi_0$, defined by $\wp(\xi_0)=2\g-1$. In order to regularize this integral we introduce a cut-off at $\xi=\xi'$, defined in the following way:
\begin{equation}
h(\xi')=\frac{1}{\e^2},
\end{equation}
solving for $\xi'$ one gets:
\begin{equation} \xi'=\wp^{-1}\left(\wp(\xi_0)+\frac{\g\e^2(4\g-3)}{1-\e^2}\right).
\end{equation}
Expanding in $\e$ we get:
\begin{equation}
 \xi'=\xi_0-\frac{\sqrt{\g}\e^2}{2}.
\end{equation}
At this point we perform the integration over $\xi$ we get:
\begin{equation}
\int_{0}^{\xi'}d\xi\frac{2 h(\xi)^2}{\sqrt{\g}}=\frac{1}{\e^2}+\log\left(\frac{2}{\e}\right)+\mathcal{B}+\mathcal{O}(\e),
\end{equation}
where  $\mc B$ has been defined in equation (\ref{Ci}).
and we have introduced the Weierstrass $\zeta$ and $\sigma$ functions. For the $Z$ integral we place a cut-off at $Z=\d$ we finally obtain
\begin{equation}
S(\g)=\frac{\Vol(\mathbb{S}^1)\Vol(\mathbb{S}^5) L^8}{4 G_N}\left(\frac{R}{\d}-1\right)\left(\frac{1}{\e^2}+\log\left(\frac{2}{\e}\right)+\mathcal{B}\right).
\end{equation} 
The holographic entanglement entropy for the vacuum  is found by considering $\g=1$:
\begin{equation}
S(\g=1)=\frac{\Vol(\mathbb{S}^1)\Vol(\mathbb{S}^5) L^8}{4 G_N}\left(\frac{R}{\d}-1\right)\left(\frac{1}{\e^2}+\log\left(\frac{2}{\e}\right)-\frac{1}{2}\right).
\end{equation}
After vacuum subtraction we obtain:
\begin{equation}
\Delta S=\frac{\Vol(\mathbb{S}^1)\Vol(\mathbb{S}^5) L^8}{4 G_N}\left(\frac{R}{\d}-1\right) \left(\mathcal{B}+\frac{1}{2}\right).
\end{equation}
The universal contribution is given by:
\begin{equation}
\Delta S_{\text{UNIV}}=-\frac{\Vol(\mathbb{S}^1)\Vol(\mathbb{S}^5) L^8}{4 G_N}\left(\mathcal{B}+\frac{1}{2}\right).
\end{equation}
\\

Notice that we get the same result independently of the regularization procedure adopted.  Also in this case our result matches the expression found in literature using the Fefferman-Graham regularization \cite{Estes:2014hka}.  

\section{Two dimensional holographic interfaces}\label{3dim}
\setcounter{equation}{0}

In this section we are going to focus on gravity solutions representing a  two dimensional interface.  It has been observed in various contexts that  in a three dimensional CFT with a  two dimensional conformal defect one can associate an effective  central charge to the defect \cite{Nozaki:2012qd,Jensen:2015swa,Gentle:2015ruo}. This central charge  appears  both in the entanglement entropy and in the Weyl-anomaly of the theory.

The fact that we can identify an effective central charge can be understood holographically. The argument is that when a 1+1 dimensional interface enjoys conformal symmetry we expect the dual bulk geometry to present an $AdS_3$ factor, we can thus associate an effective central charge to the interface through the Brown-Henneaux formula \cite{BH}. This was first done in \cite{Gentle:2015ruo} in the context of type IIB supergravity solutions dual to half-BPS disorder-type surface defects in $\mathcal{N}=4$ Super Yang-Mills theory. It was also observed that the effective central charge arising from the Brown-Henneaux formula was the same quantity that appears in the computation of the entanglement entropy. In this section we explore other examples of a 1+1 dimensional interface which enjoys conformal symmetry.
	
In particular we focus on examples where the 1+1 interface is embedded in a 3 dimensional theory. In addition to the computation of entanglement entropy we calculate the conformal anomaly and show that it is governed by the same central charge appearing in the entanglement entropy computation and arising from the Brown-Henneaux formula. Before going over explicit examples we prove the following statement: in an ICFT with an even dimensional interface embedded into an odd dimensional spacetime the universal contribution of entanglement entropy for a spherical entangling surface centered on the interface is equal to minus the universal term of free energy on a sphere. 

We explicitly prove this statement for a 3 dimensional theory with a 2 dimensional interface. The generalization to arbitrary dimensions is straightforward. The proof follows closely section 4 of \cite{Casini:2011kv}. The field theory lives on a three dimensional spacetime given by:
\begin{equation}
ds^2=-dt^2+d\r^2+\r^2 d\phi^2,
\end{equation}
where we have chosen polar coordinate for the spatial slice. The interface is located at $\sin \phi=0$. We perform the following change of coordinates:
\begin{eqnarray}
t&=&\frac{R \cos \eta \sinh(\tau/R)}{1+\cos\eta \cosh(\tau/R)}\nonumber\\
\r&=&R \frac{\sin \eta}{1+\cos \eta \cosh(\tau/R)}.
\end{eqnarray}
The spacetime is then given  by
\begin{eqnarray}
ds^2&=&\Omega^2(-\cos^2 \eta d\tau^2+R^2(d\eta^2+\sin^2 \eta d\phi^2)) \nonumber\\
\Omega&=&(1+\cos^2\eta \cosh(\tau/R))^{-1},
\end{eqnarray}
which, after removing $\Omega$, corresponds to the static patch of de Sitter space with curvature scale R. It can be shown (for details see \cite{Casini:2011kv}) that the new coordinates cover the causal development of the ball $\rho<R$ on the surface $t=0$ (which is exactly our entangling region). In addition one can show that the modular flow generated by the modular Hamiltonian in the causal diamond corresponds to time flow in this new coordinate system and that original density matrix can be written as a thermal density matrix with temperature $T=1/(2\pi R)$. This implies that the entanglement entropy of the ball shaped region can be written as a thermal entropy:
\begin{equation}\label{thermo}
S=\beta E-W,
\end{equation}
where $W$ is the free energy and $E$ is the expectation value of the operator which generates time evolution, explicitly:
\begin{equation}
E=\int_V d^{2} x \sqrt{h} \braket{T_{\mu \nu}} \xi^\mu n^\nu=-\int_V d^{2} x  \sqrt{-g} \braket{T^\tau _{~\tau}},
\end{equation}
where $V$ is a constant $\tau$ slice, $n$ is the unit normal $n^\mu \partial_\mu=\sqrt{|g_{\tau \tau}|}\partial_\tau$ and $\xi$ is the Killing vector that generates $\tau$ translations $\xi^\mu\partial_\mu=\partial_\tau$.\\
To compute $E$ we need to write an expression for $\braket{T^\tau _{~\tau}}$. A powerful tool to do that is symmetry. In fact we know that the interface is extended along the surface $\sin \phi=0$ which corresponds to a two dimensional de Sitter spacetime. The isometry of de Sitter space forces the stress tensor to satisfy the following relations:
\begin{eqnarray}
\braket{T^\alpha_{~\beta}}&=&\tilde c \; \delta^\alpha_{\; ~\beta}\; \delta(\sin \phi) \nonumber \\
\braket{T^\phi_{~\beta}}&=&\braket{T^\alpha_{~\phi}}=\braket{T^\phi_{~\phi}}=0,
\end{eqnarray}
where $\alpha$ and $\beta$ denote any of the coordinates $\eta$ and $\tau$. This suffices to show that $E$ is finite. On the other side, since the interface is even dimensional we expect a logarithmic divergence in both $S$ and $W$. This means that $E$ does not contribute to the universal terms in equation (\ref{thermo}), thus:
\begin{equation}
S_{\text{UNIV}}=-W_{\text{UNIV}}.
\end{equation}
In order to find $W_{\text{UNIV}}$ we go to imaginary time with periodicity $2 \pi R$. The metric becomes
\begin{equation}
ds^2=\cos^2 \theta d\tau^2+R^2(d\theta^2+\sin^2 \theta d\phi^2),
\end{equation}
which we recognize as the metric of $\mathbb S^3$ once we identify $\tau \sim \tau+2 \pi R$. Thus:
\begin{equation}\label{eq}
S_{\text{UNIV}}=-W_{\text{UNIV}}(\mathbb S^3),
\end{equation}
as anticipated. 

We would like to relate this quantity to an effective central charge (since we are in presence of a two dimensional conformal field theory living on the interface). To do that we focus on $W_{\text{UNIV}}(\mathbb S^3)$. For definiteness let's say we locate the interface at the equator of the sphere. By the same symmetry arguments as in the de Sitter case we have:
\begin{eqnarray}
\braket{ T_{\vartheta \vartheta}}&=&\braket{ T_{\vartheta \alpha}}=0 \nonumber \\
\braket{ T_{\alpha \beta}}&=&\frac{c_{\text{eff}}}{24 \pi r^2} h_{\alpha \beta} \delta \left(\vartheta-\frac{\pi}{2}\right), \label{T}
\end{eqnarray}
where $\alpha$ and $\beta$ denotes the directions along the interface and $h$ is the metric of the sphere
 \begin{equation} \label{metricsphere}
 ds^2_{\mathbb{S}_3}=r^2 \left(d\vartheta +\sin^2 \vartheta ds^2_{\mathbb{S}^2}\right),
 \end{equation}
 with $\vartheta \in [0,\pi]$ and $\vartheta=\pi/2$ corresponding to the location of the interface. If we change the radius of the sphere by $\d r$ we have:
\begin{equation}
\delta_r  W_{\text{UNIV}}=\frac{1}{2} \int_{\mathbb S^3} d^3x \sqrt{h} \d h^{i j} \braket{ T_{ij}}=-\frac{c_{\text{eff}}}{3r} \d r=-\frac{c_{\text{eff}}}{3}\d_r \log r,
\end{equation}
where we have used equations (\ref{T}) to get the final result. This shows that the coefficient of the logarithmic term of entanglement entropy is related to the coefficient of the Ricci scalar in the conformal anomaly\footnote{If the interface is even dimensional embedded into a odd dimensional spacetime of general dimension we have that the coefficient of the logarithmic term is related to the A anomaly.}. 

Notice that a priori this is a non trivial fact. In a two dimensional CFT the only central charge is the coefficient of the Ricci scalar in the trace anomaly, but in a ICFT the situation is more complicated. In fact the 1+1 dimensional interface is embedded in a higher dimensional spacetime where the theory lives, thus other terms, such as the trace of the extrinsic curvature, could contribute to the trace anomaly. 

In the following we are going to focus on specific examples. We are going to compute both entanglement entropy and free energy holographically and we will show that equation (\ref{eq}) holds. To find the free energy holographically write the metric in the same form as in equation (\ref{metric}), replacing $AdS_3$ with its Euclidean counterpart, named $H_3$
\begin{equation}
ds^2_{H_3}=\frac{1}{\cos^2 \t} \left(d \t^2+\sin^2\t ds^2_{\mathbb S_2}\right)
\end{equation}
where $\t \in [0, \pi/2]$ and we have sliced $H_3$ using spheres. The free energy can then be computed holographically as the on shell action $I_{\text{on shell}}$. We are going to use ony the double cut off procedure, one can obtain the same results using the single cut off regulator.

\subsection{3 dimensional Einstein-Dilaton Janus}\label{3dimDE}
The first example we discuss is a bottom up system. We can construct an ICFT from a CFT by considering a marginal operator $\mc O$ and assigning to it a coupling constant that jumps across a 1+1 dimensional plane. We construct the bulk theory dual to this deformation by solving the equations of motion derived from the action $I$ of a massless field $\Phi$, dual to $\mc O$, minimally coupled to the metric. In particular one has
\begin{equation}
I=\frac{1}{16 \pi G_N}\int d^{4}x\sqrt{-g}\left(R-\partial_\mu \Phi \partial^\mu \Phi +\frac{6}{L^2}\right),
\end{equation}
from which one finds:
\begin{eqnarray}
ds_{\pm}^2&=&\frac{L^2}{q_{\pm}^2}\left(\frac{dq_{\pm}^2}{P(q_\pm)}+ds^2_{AdS_{3}}\right) \nonumber \\
\Phi(q_{\pm})&=&\Phi_0\pm \lambda \int_{q_{\pm}}^{q_{*}}\frac{x^{2}}{\sqrt{P(x)}}dx, \label{metricjanus3d} 
\end{eqnarray}
where $P(x)=1-x^2+\frac{\lambda^2}{6}x^{6}$ and $q_{*}$ is defined by $P(q_{*})=0$. The parameter $\l$ quantifies the strength of the Janus deformation, $\lambda \in [0,2 \sqrt{2}/3]$ and one recovers $AdS_{4}$ for $\lambda=0$. Notice that the bulk geometry is covered using two different patches, the patches smoothly join at $q_\pm=q_{*}$ while the boundary is located at $q_{\pm}=0$. There are two boundary regions (glued together at $Z\rightarrow 0$) that correspond to the two different sides of the interface.

\subsubsection{Holographic Entanglement Entropy}
As usual we take the entangling region to be a ball or radius $R$ centered on the interface. From equation (\ref{HEE}) we get:
 \begin{equation}
 S=\frac{4 L^{2}}{4 G_N} \int_{\d}^{R}\frac{R (R^2-Z^2)^{-\frac{1}{2} }dZ}{Z} \int_{\e}^{q_*}\frac{d q}{q^{2}\sqrt{P(q)}}.
 \end{equation}
Working with the double cut-off regulator requires to compute
 \begin{equation}
 \mc I= \int_{\e}^{q_*}\frac{d q}{q^{2}\sqrt{P(q)}}.
 \end{equation}
 The expression of $q_*$ as a function of $\l$ is:
 \begin{equation}
 q_*^2=-2 \sqrt{2} \sqrt{\frac{1}{\lambda ^2}} \cos \left(\frac{1}{3} \left(2 \pi -\tan ^{-1}\left(\sqrt{\frac{8}{9 \lambda ^2}-1}\right)\right)\right).
 \end{equation}
 We change variable of integration by introducing $t=q/q_*$:
 \begin{equation}
 \int_{\e/q_*}^{1}\frac{d t}{q_*t^{2}\sqrt{P(q_* t)}}.
 \end{equation}
 Using the fact that $q_*^2=1+\frac{\l^2}{6}q_*^6$ one can write
 \begin{eqnarray}
 P(q_* t)&=&(1-t^2)\left(1-\frac{\l^2 q_*^{6}t^2}{6}({t^{2}+1})\right) \label{P}\nonumber\\
 &=&\frac{\l^2 q_*^{6}}{6}(1-t^2)\left(t^2+\frac{1+\sqrt{1-\frac{24}{\l^2 q_*^6}}}{2}\right)\left(-t^2+\frac{-1+\sqrt{1-\frac{24}{\l^2 q_*^6}}}{2}\right)\nonumber\\
 &=&\frac{\l^2 q_*^{6}}{6}(b-t^2)(t^2-d)(a-t^2),
 \end{eqnarray}
 where $b=1$, $d=-\frac{1+\sqrt{1-\frac{24}{\l^2 q_*^6}}}{2}$ and $a=\frac{-1+\sqrt{1-\frac{24}{\l^2 q_*^6}}}{2}$. Using the change of coordinate $t^2=s$, we write the integral in the following form:
 \begin{equation}
 \frac{\sqrt{6}}{2 q_*^4 \l} \int_{u}^{b} \frac{ds}{s \sqrt{(s-c)(b-s)(s-d)(a-s)}},
 \end{equation}
 with $c=0$ and $u=(\e/q_*)^2$. We note that for $\lambda \in [0, 2 \sqrt{2}/3]$ we have $d<c<u<b<a$. This is an elliptic integral and can be found in \cite{GR}. It evaluates to:
 \begin{eqnarray}
 \mc I &=&\frac{\sqrt{6} \left((a-b) \Pi \left(\frac{a(b-c)}{b(a-c)};\chi \left|k^2\right.\right)+b F\left(\chi \left|k^2\right.\right)\right)}{\left( q_*^4 \l \right) \left(a b \sqrt{(a-c) (b-d)}\right)}\label{int}\nonumber\\
 \chi &=&\sin ^{-1}\left(\sqrt{\frac{(a-c) (b-u)}{(a-u) (b-c)}}\right)\nonumber\\
 k&=&\sqrt{\frac{(a-d) (b-c)}{(a-c) (b-d)}}.
 \end{eqnarray}
 We expand (\ref{int}) for small $\e$, we get:
 \begin{eqnarray}
 \mc I&=&\frac{1}{\e}+C(\l)+\mc O(\e)\nonumber\\
 C(\l)&=&\frac{\sqrt{6} (a-a d)^{3/2} \left((d-1) E\left(\frac{a-d}{a-a d}\right)-d K\left(\frac{a-d}{a-a d}\right)\right)}{d \lambda  (a-d)^2}.
 \end{eqnarray}
 $K(x)$ and $E(x)$ denote the complete elliptic integral of first and second kind.
 \begin{figure}\centering
 	\includegraphics[width=10 cm]{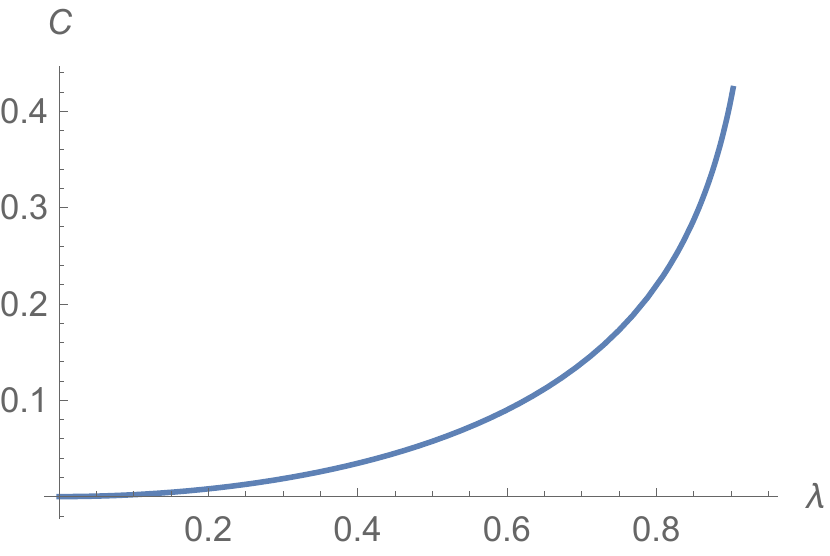}
 	\caption{Behavior of $C(\l)$}
 	\label{C}
 \end{figure}
 The divergent term is $\l$ independent, so taking into account that
 \begin{equation}
 \int_{\d}^{R}\frac{R dZ}{Z \sqrt{R^2-Z^2}}=-\log\left(\frac{\d/R}{1+\sqrt{1-\d/R}}\right)\approx \log\left(\frac{2R}{\d}\right)+\mc O(\d)
 \end{equation}
 and subtracting the background contribution we are left with:
 \begin{equation}
 \Delta S =\frac{L^2}{G_N} C(\l) \log\left(\frac{2R}{\d}\right).
 \end{equation}
 It is natural to identify an effective central charge as 
 \begin{equation}\label{c1}
 c_{\text{eff}}=  \frac{3 L^2}{G_N} C(\l).
 \end{equation}
 The behavior of $C(\l)$ is displayed in figure \ref{C}. 
  
 \subsubsection{On shell Action} 
 The Euclidean action of the bulk theory is given by:
 \begin{equation}\label{action}
 I=-\frac{1}{16 \pi G_N} \int_{M_4} \left(R- g^{ab} \partial_a \Phi \partial_b \Phi +\frac{6}{L^2}\right)-\frac{1}{8 \pi G_N} \oint_{\partial M_4} K,
 \end{equation}
 where the second term is the Gibbons-Hawking contribution, added to ensure a sensible variational principle. It is then natural to decompose the on shell action into two contributions, one coming from the Gibbons-Hawking term and the other one coming from the bulk integration. Using Einstein equation we have:
 \begin{eqnarray}
 I_{\text{on shell}}&=&I_{\text{bulk}}+I_{\text{surface}}\nonumber\\
 I_{\text{bulk}}&=& \frac{3}{8 \pi G_N L^2} \int_{M_4} d^4 x \sqrt{g}\nonumber\\
 I_{\text{surface}}&=&-\frac{1}{8 \pi G_N} \oint_{\partial M_4} K.
 \end{eqnarray}  
 It is easy to show that $I_{\text{surface}}$ does not contain any logarithmic divergences. We focus only on $I_{\text{bulk}}$. We write it as:
 \begin{equation}
 I_{\text{bulk}}=\frac{3 L^2}{G_N} \int_{0}^{\t_0}d\t \frac{\sin^2 \t}{\cos^3 \t} \int_{\e}^{q_*}dq \frac{1}{q^4 \sqrt{P(q)}},
 \end{equation}
 where $\cos \t_0=\frac{\d}{r }$. Let's look at the $q$ integration first
 \begin{equation}
 \mc J(\e)=\int_{\e}^{q_*}dq \frac{1}{q^4 \sqrt{P(q)}}.
 \end{equation}
We expand $\mc J$ in Laurent expansion\footnote{The fact that the divergent pieces are not $\l$ dependent is showed in the following.}:
 \begin{equation}\label{J}
 \mc J(\e)=\frac{A_{-3}}{\e^3}+\frac{A_{-1}}{\e}+A(\l)+A_{1}(\l)\e+...
 \end{equation} 
 We can perform the $\t$ integration, getting:
 \begin{equation}
 I_{\text{bulk}}=\frac{3L^2}{G_N}\left(\frac{1}{2}\left(\frac{r }{\d}\right)^2+\frac{1}{2}\log\left(\frac{\d}{2 r }\right)-\frac{1}{4}\right)\left(\frac{A_{-3}}{ \e^3}+\frac{A_{-1}}{\e}+A(\l)\right).
 \end{equation} 
 The constant term in $\mc J$ gives a logarithmic divergence in the on shell action equal to 
 \begin{equation}
 \frac{3 L^2}{2 G_N}A(\l) \log\frac{\d}{r}.
 \end{equation}
 Thus we find:
 \begin{equation}\label{c2}
 c_{\text{eff}}=\frac{9L^2}{2 G_N} A(\l).
 \end{equation}
 In the following paragraph we show that this central charge matches equation (\ref{c1}) by proving that \begin{equation}\label{CeqA}
 A(\l)=\frac{2}{3} C(\l).
 \end{equation}
 This will imply that equation (\ref{eq}) is satisfied. We will also show that the divergent term in equation (\ref{J}) are $\l$ independent.
 We start as usual by changing variable $t=q/q_*$ and using equation ($\ref{P}$):
 \begin{eqnarray}
 \mc J&=& \int_{\e/q_*}^{1} \frac{dt}{q_*^3 t^4 \sqrt{P{q_* t}}}\nonumber\\
 &=& \frac{1}{q_*^3} \sum_{k=0} \frac{(1/2)_k}{k!} \left(\frac{\l^2 q_*^6}{6}\right)^k \int_{\e/q_*}^{1}  dt t^{2 k-4} (t^2+1)^k(1-t^2)^{-1/2},
 \end{eqnarray}
 where we have expanded $P(q_* t)^{-1/2}$ in series. We are interested in the divergent terms and in the constant term. The divergent terms come from $k=0,1$, while to find the $\e$ independent contribution of this integral, we can simply evaluate the primitive of the integrand at $t=1$. One gets:
 \begin{eqnarray}
 A_{-3}&=&\frac{1}{3 }\nonumber\\
 A_{-1}&=&\frac{1}{2 q_*^2 }\left(1+\frac{\l^2 q_*^6}{6}\right)=\frac{1}{2 }\nonumber\\
 A(\l)&=&\frac{1}{q_*^3} \sum_{k=0} \frac{(1/2)_k}{k!} \left(\frac{\l^2 q_*^6}{6}\right)^k a_k\nonumber\\
 a_k&=&-\frac{1}{2} \sqrt{\pi } \left(k \, _2\tilde{F}_1\left(1-k,k-\frac{3}{2};k;-1\right)+(1-2 k) \, _2\tilde{F}_1\left(k-\frac{3}{2},-k;k;-1\right)\right) \Gamma \left(k-\frac{3}{2}\right),\nonumber
 \end{eqnarray}
 where $ _2\tilde{F}_1(a,b;c;z)$ is the regularized hypergeometric function $_2 F_1(a,b;c;z)/\Gamma(c)$. Using the same technique with $\mc I$ one gets the following expression for $C(\l)$:
 \begin{eqnarray}
 C(\l)&=&\frac{1}{q_*} \sum_{k=0} \frac{(1/2)_k}{k!} \left(\frac{\l^2 q_*^6}{6}\right)^k c_k\\
 c_k&=&-\frac{1}{2} \sqrt{\pi } k \left(\, _2\tilde{F}_1\left(1-k,k-\frac{1}{2};k+1;-1\right)-2 \, _2\tilde{F}_1\left(k-\frac{1}{2},-k;k+1;-1\right)\right) \Gamma \left(k-\frac{1}{2}\right).\nonumber
 \end{eqnarray}
 By using the fact that $q_*^2=1+\frac{\l^2 q_*^6}{6}$ and that $a_0=c_0=0$ we can write:
 \begin{eqnarray}
 A(\l) q_*^3&=&\sum_{k=1} \frac{(1/2)_k}{k!} \left(\frac{\l^2 q_*^6}{6}\right)^k a_k\nonumber\\
 C(\l) q_*^3&=&\sum_{k=1} \frac{(1/2)_k}{k!} \left(\frac{\l^2 q_*^6}{6}\right)^k \left(c_k+\frac{(1/2)_{k-1}}{(1/2)_k}k c_{k-1}\right).
 \end{eqnarray}
 Using the properties of hypergeometric functions and gamma function one notes that
 \begin{equation}
 c_k+\frac{(1/2)_{k-1}}{(1/2)_k}k c_{k-1}=\frac{3}{2} a_k
 \end{equation}
 which proves equation (\ref{CeqA}).
 
Summing up we have:
 \begin{equation}
 I_{\text{bulk}}=\frac{3L^2}{G_N}\left(\frac{1}{2}\left(\frac{r }{\d}\right)^2+\frac{1}{2}\log\left(\frac{\d}{2 r }\right)-\frac{1}{4}\right)\left(\frac{1}{3 \e^3}+\frac{1}{2 \e}+A(\l)\right),
 \end{equation} 
 once we subtract the vacuum contribution we get:
 \begin{equation}
 \Delta I_{\text{bulk}}=\frac{3L^2 A(\l)}{G_N}\left(\frac{1}{2}\left(\frac{r }{\d}\right)^2+\frac{1}{2}\log\left(\frac{\d}{2 r }\right)-\frac{1}{4}\right).
 \end{equation}
 Notice that the fact that the divergent terms in (\ref{J}) are $\l$ independent makes the final result depending only on the interface cut-off $\d$.  
 \subsubsection{Brown-Henneaux fomula}
 One last check we can perform is whether our effective charge could be derived from Brown-Henneaux formula:
 \begin{equation}
 c_{\text{eff}}=\frac{3 L}{2 G_N^{(3)}}.
 \end{equation}
 Of course the gravitational constant $G_N$ that has appeared so far is a 4 dimensional Newton constant. In order to obtain the three dimensional counterpart we reduce on the non compact direction $q$. In order to obtain a finite result we subtract the vacuum contribution, from a more physical point of view this is done to isolate the interface contribution. Note that we have to take into account the non trivial $q$-dependent factor that appears in front of the $AdS_3$ space in the metric (\ref{metricjanus3d}). In particular we have:
 \begin{eqnarray}
 \frac{1}{G_N^{3}}&=&\frac{2}{G_N}\Delta\left( \int \frac{L dq}{q^2 \sqrt{P(q)}}\right)\\
 &=&  \frac{2 C(\l) L}{G_N},
 \end{eqnarray}
 where we have used the results derived in the computation of the holographic entanglement entropy. 
 Using Brown-Henneaux formula we then have:
 \begin{equation}
 c_{\text{eff}}=\frac{3 L^2}{G_N} C(\l),
 \end{equation}
 which agrees with effective central charge obtained in   (\ref{c1}).

\subsection{M-theory Janus}
The M-theory Janus solution is a one parameter deformation of the $AdS_4 \times \mathbb{S}^7$ vacuum solution of the eleven dimensional supergravity \cite{D'Hoker:2009gg}. The dual field theory is ABJM theory deformed by a primary operator of dimension two localized on a interface.

The bulk metric is given by
\begin{equation}
ds^2=f_1^2 g_{AdS_3}+f_2^2 g_{\mathbb{S}^3_2}+f_3^2 g_{\mathbb{S}^3_3}+4 \rho^2(dx^2+dy^2).
\end{equation}
where all the functions appearing in the metric depend on the coordinates $x$ and $y$ and on a parameter $\lambda$. The coordinates $x,y$ parametrize a strip, while the deformation parameter is real and one recovers pure $AdS_4 \times \mathbb{S}^7$ for $\l=0$. In particular one has:
\begin{eqnarray}
f_1&=&\frac{\cosh(2 x)}{\sqrt{1+\l^2}} \nonumber F_+(x,y)^{1/6}F_-(x,y)^{1/6} \nonumber \\
f_2&=& 2 \cos(y) F_+(x,y)^{1/6}F_-(x,y)^{-1/3} \nonumber \\
f_3&=& 2 \sin(y) F_+(x,y)^{-1/3}F_-(x,y)^{1/6}\nonumber \\
\rho&=& F_+(x,y)^{1/6}F_-(x,y)^{1/6}\nonumber \\
F_+(x,y)&=& 1+ 2\l (\sinh(2x)+\l)\cos^2(y)/\cosh^2(2x)\nonumber\\
F_-(x,y)&=& 1- 2\l (\sinh(2x)-\l)\sin^2(y)/\cosh^2(2x).
\end{eqnarray}
\subsubsection{Holographic Entanglement Entropy}
We can use equation (\ref{HEE}) to find the entanglement entropy of a spherical region centered on the interface, we find:
\begin{eqnarray}
S&=&\frac{ 2 \Vol(\mathbb{S}^3)^2}{4 G_N} \int_{\text{cut-off}}^{R} \frac{R (R^2-Z^2)^{-\frac{1}{2} }dZ}{Z} \int dx dy 2 f_1 f_2^3 f_3^3 \rho^2\\
&=&\frac{ 2 \Vol(\mathbb{S}^3)^2}{4 G_N} \int_{\text{cut-off}}^{R} \frac{R (R^2-Z^2)^{-\frac{1}{2} }dZ}{Z} \int_{0}^{\pi/2} dy (2 \sin (2 y))^3 \int_{-\text{cut-off}}^{\text{cut-off}} dx\frac{\cosh(2x)}{\sqrt{1+\l^2}} \nonumber
\end{eqnarray}
We start by using the two cut-off procedure. We place two independent cut-off, one for the $Z$ integral located at $Z=\d$, the other for the $x$ integral, located at $f_1=1/\e$, i.e $x=x_\infty(\l)=1/2 \cosh^{-1}(\sqrt{1+\l^2}\e)$. This procedure gives $\Delta \mc A=0$. That is because
\begin{eqnarray}\label{Delta}
\begin{split}
\Delta \int \frac{\cosh(2x)}{\sqrt{1+\l^2}}&=&2\Delta\int_{0}^{x_{\infty}(\l)}\frac{\cosh(2x)}{\sqrt{1+\l^2}}dx \\
&=&\Delta \frac{\sinh(2 x_\infty(\l))}{\sqrt{1+\l^2}}\xrightarrow{\lim \e\rightarrow 0}0
\end{split}
\end{eqnarray}  In this case it is interesting to look also at the single cut-off procedure. We place the cut-off at
\begin{equation}\label{cutoffMjanus}
f_1/Z=\frac{1}{\d}.
\end{equation}
Since we still have to integrate over $Z$ (and the lower bound of the $Z$ integral is linear in $\d$) we cannot assume $Z/\d<<1$. However we can still express the solution of (\ref{cutoffMjanus}) as a Laurent series with respect to $\d/Z$. In particular we have:
\begin{equation}
\cosh(2 x_\infty)=\frac{Z \sqrt{1+\l^2}}{\d}\left(1+\sum_{k \text{ even}}c_k(y) \left(\frac{\d}{Z}\right)^k\right),
\end{equation}
where the summation index $k$ is a positive even number.
The $x$ integration can now be carried out, we obtain:
\begin{equation}
\int_{-x_\infty}^{x_\infty} dx\frac{\cosh(2x)}{\sqrt{1+\l^2}}=\frac{Z}{ \d} \left(1+\sum_{k \text{ even}} \tilde c_k (y) \frac{\d}{Z} \right)^k,
\end{equation}
for some coefficients $\tilde c_k(y)$. We now proceed to the $Z$ integration, we notice that:
\begin{eqnarray}
\int_{\d/\sqrt{1+\l^2}}^{R} \frac{R (R^2-Z^2)^{-\frac{1}{2} }dZ}{\d}&=&\frac{\pi R}{2 \d}+\frac{1}{\sqrt{1+\l^2}}+\mathcal{O(\d)}\nonumber\\
\d^k  \int_{\d/\sqrt{1+\l^2}}^{R} \frac{R (R^2-Z^2)^{-\frac{1}{2} }dZ}{Z^k}&\approx&\mc O(\d),
\end{eqnarray}
where $\mc O (\d)$ denotes linear and higher orders in $\d$.
Thus, neglecting all term $\mc O (\d)$, we finally get:
\begin{equation}\label{minA}
S= \frac{8}{3 G_N} \Vol(\mathbb{S}^3)^2\left(\frac{\pi R}{\d}+\frac{1}{\sqrt{1+\l^2}}\right).
\end{equation}
Note that, as in the two cut-off scheme, we don't obtain  a logarithmic term. One might be worried that the constant term is different in the two schemes, however it has been shown before  that the constant appearing in the computation of holographic entanglement entropy for this set up is not a universal quantity \cite{Estes:2014hka}. Hence different regularization schemes determine that the effective central charge vanishes, i.e.  $c_{\text{eff}}=0$.

\subsubsection{On shell Action}
We look at the Euclidean on shell action for M-theory Janus. Since we want to place the dual CFT on a sphere we choose global coordinates for the  AdS  factor. The Euclidean action is given by:
\begin{equation}
I=-\frac{1}{2 \kappa_{11}^2}\int d^{11} \sqrt{g} \left(R-\frac{1}{48}F_{MNPQ}F^{MNPQ}\right)-\frac{i}{12 \kappa_{11}^2}\int C \wedge F\wedge F
\end{equation}
where $C$ is a 3-form potential and $F=d C$. Using the equation of motion we have $R=\frac{1}{144} F_{MNPQ}F^{MNPQ}$. We then have:
\begin{eqnarray}
2 \kappa_{11}^2 I_{\text{on shell}}=\frac{1}{72}\int_{\mc M_\d} d^{11}x  \sqrt{g} F_{MNPQ}F^{MNPQ}+\frac{1}{6} \int_{\mc M_{\d}} C \wedge F \wedge F+ 2 \kappa_{11}^2 I_{\text{GH}}.
\end{eqnarray}
Notice that we have introduced a cut-off $\d$ (we will be more precise about it later), the regularized manifold has been named  $\mc M_\d$. Furthermore we have included the Gibbons Hawing term. Let's focus on the first two term first. By writing $\sqrt{g}F_{MNPQ}F^{MNPQ}$ as $4! (*F \wedge F)$, $F=dC$ and integrating by parts one gets:
\begin{equation}
2 \kappa_{11}^2 I_{\text{on shell}}= -\frac{1}{3}\left(\int_{\mc M_\d} d\left(*F \wedge C\right)+\frac{1}{2}\int_{\mc M_\d} d\left(C \wedge F \wedge C\right)\right)+ 2 \kappa_{11}^2 I_{\text{GH}},
\end{equation}
where we have omitted the terms that vanish due to the equations of motion. We can now use Stokes theorem to express this integral as a boundary term. Notice also that since $F$ is a 4-form and $C$ is a 3-form we have $C\wedge F\wedge C= F \wedge C\wedge C=-F \wedge C\wedge C=0$. Thus we have:
\begin{equation}
2 \kappa_{11}^2 I_{\text{on shell}}= -\frac{1}{3}\left(\int_{\partial \mc M_\d} *F \wedge C\right)+ 2 \kappa_{11}^2 I_{\text{GH}}.
\end{equation}
We work in the two cut-offs scheme. This means that we place a cut-off at $\theta=\theta_0=\arccos \d$ and we compute all quantities with respect to vacuum solution. The $\d$ cut-off is a natural physical cut-off for the interface, of course generally speaking before the vacuum subtraction we have another source of divergence (coming from the $x$ integration), we then introduce a cut-off also at large $x$. That cut-off is not physical since will be removed by the vacuum subtraction.

The expression for $C$ in our set up is given by:
\begin{equation}
C=b_1(x,y) \hat \omega_{AdS_3}+b_2(x,y) \hat \omega_{\mathbb{S}^3_2}+b_3(x,y) \hat \omega_{\mathbb{S}^3_3},
\end{equation}
the $\hat \omega$'s are the volume forms of the $AdS$ space and 3-spheres with unit radii. Notice that since the boundary is $\cos \theta=\d$ the only non zero term is of the form $ e^1 \wedge e^2 \wedge e^3 ... \wedge e^{11}$ (the index $0$ refer to the coordinate $\theta$), however that term does not appear in $*F\wedge C$. This means
\begin{equation}
I_{\text{on shell}}=I_{\text{GH}}.
\end{equation}
Using the explicit solution of \cite{D'Hoker:2009gg} we find that:
\begin{equation}
\int_{\partial \mc M_\d}\sqrt{\gamma} K=\frac{\sin \theta_{0}}{\cos^2 \theta_0} \Vol(\mathbb{S}^2)\Vol(\mathbb{S}^3)^2\frac{128}{\sqrt{1+\lambda^2}} \int dx dy \cosh x \sin^3 y.
\end{equation}
Notice that the x integral is divergent this is because we are working in the two cut-off scheme and we should always perform a vacuum subtraction before declaring a quantity physical. Subtracting the vacuum contribution gives and using that $\cos \theta_0=\d$:
\begin{equation}
\Delta \int_{\partial \mc M_\d}\sqrt{\gamma} K=0 .
\end{equation}
It is clear that there isn't any logarithmic divergence in $\d$. This means that the on shell action does not change as we vary the radius of the sphere where the CFT lives, i.e. $c_{eff}=0$.

\subsubsection{Brown-Henneaux formula}
The fact that $c_{\text{eff}}$ is zero can be understood also using Brown Henneaux formula:
\begin{equation}
c_{\text{eff}}=\frac{2}{3 G_N^{11}}\Delta\int dx dy f_1 f_2^3 f_3^3  \rho^2=0.
\end{equation}
Where we have used the same technique used in equation (\ref{Delta}). Notice that the fact that the effective charge is zero does not imply the absence of conformal anomaly in general. In fact the interface is embedded in a higher dimensional space, this means that one can make scalar quantities (such as the trace of the extrinsic curvature) which can contribute to the conformal anomaly.

\section{Conclusions}
\setcounter{equation}{0}
In this paper we have presented a new cut-off procedure (called ``double cut-off" regularization) that can be used to regularize divergent bulk quantities in holographic spacetimes which realize interface CFTs.

The motivation for this cut-off procedure relies on the fact that a d-dimensional conformal field theory with a $d-1 $ dimensional interface, has a bulk dual can be constructed using a warped spacetime an $AdS_d$ factor . This choice of coordinates makes manifest the symmetry group that characterizes the set up. In particular it is natural to regard the $AdS$ slices as dual to the interface, since they share the same symmetry. There is then a natural bulk cut-off realized by limiting the holographic coordinate of the $AdS$ slice. We expect this cut-off procedure to be well defined only when computing quantities that are intrinsic to the interface. A physical quantity can be made intrinsic by subtracting the vacuum contribution. To make this quantity finite before the vacuum subtraction we need to introduce a second cut-off which we consider as a mere tool for  intermediate steps.

We tested this procedure for set ups where the holographic entanglement entropy is already known, finding agreement with the results available in the existing literature \cite{Estes:2014hka}. Of particular interest is the case of $1+1$ dimensional interfaces. In that case it is natural to associate a central charge to the set up through the Brown-Henneaux formula. We verified that this effective central charge plays the role one would naively expect in the computation of entanglement entropy and conformal anomaly.

We stress that the main advantage of the double cut-off regularization procedure is to simplify considerably the computations one needs to perform to calculate any quantity (such as entanglement entropy and on shell action) on the bulk side. This provides a new method to explore more complicated  solutions  that have been beyond reach due to the lack of Fefferman-Graham coordinates. Examples of such solutions are multi Janus solution \cite{D'Hoker:2007xz,Chiodaroli:2011nr}, which correspond to junctions of several CFTs.

\smallskip

\section*{Acknowledgements}

The work reported in this note   is supported in part by the National Science Foundation under grant  PHY-16-19926.  

\newpage

\appendix

\section{Appendix: Einstein-Dilaton Janus}\label{dilatoneinsteinjanus}
\setcounter{equation}{0}

In this appendix we discuss the $d$ dimensional generalization of the Einstein-Dilaton system studied in section \ref{3dimDE}. The expressions for the fields are given by \cite{Bak:2015jxd}:
\begin{eqnarray}
ds_{\pm}^2&=&\frac{L^2}{q_{\pm}^2}\left(\frac{dq_{\pm}^2}{P(q_\pm)}+ds^2_{AdS_{d}}\right) \label{metricjanus}\nonumber\\
\Phi(q_{\pm})&=&\Phi_0\pm \lambda \int_{q_{\pm}}^{q_{*}}\frac{x^{d-1}}{\sqrt{P(x)}}dx, \label{dilatonjanus}
\end{eqnarray}
where $P(x)=1-x^2+\frac{\lambda^2}{d(d-1)}x^{2d}$ and $q_{*}$ is defined by $P(q_{*})=0$. As in the 3 dimensional case the parameter $\l$ quantifies the strength of the Janus deformation, $\lambda \in [0,\sqrt{d-1}\left(\frac{d-1}{d}\right)^\frac{d-1}{2}]$ and one recovers $AdS_{d+1}$ for $\lambda=0$. The bulk geometry is covered using two different patches that smoothly join at $q_\pm=q_{*}$ while the boundary is located at $q_{\pm}=0$. There are two boundary regions (glued together at $Z\rightarrow 0$) that correspond to the two different sides of the interface.

Choosing as usual the entangling surface to be a sphere of radius $R$ one can use equation (\ref{HEE}) to write the following expression for the entanglement entropy: 
\begin{equation}
S=\frac{2 L^{d-1} \Vol(\mathbb{S}^{d-3})}{4 G_N}\int_{\text{cut-off}}^{R}\frac{R (R^2-Z^2)^{\frac{d-4}{2} }dZ}{Z^{d-2}} \int_{\text{cut-off}}^{q_*}\frac{d q}{q^{d-1}\sqrt{P(q)}}.
\end{equation}
We are going to study this expression up to second order in the Janus deformation parameter $\l$. Independently of the cut-off procedure we will choose we start by performing the $q$ integral:
\begin{equation}
\mathcal{P}_3(\e, d)\equiv \int_{\e}^{q_*}\frac{d q}{q^{d-1}\sqrt{P(q)}}
\end{equation} 
where $\e=\d/Z$ in the case of the single cut-off regularization, while $\e$ is simply a constant for the double cut-off regularization.

To order $\l^2$ we find
\begin{equation}
q_*=1+\frac{\l^2}{2 d (d-1)}.
\end{equation}
In order to work perturbatively in $\l$ we change variable of integration by defining $t=q/q_*$. We then write
\begin{equation}
\mathcal{P}_3(\e,d)=\int_{\e/q_*}^{1}\frac{dt}{q_*^{d-2} t^{d-1} \sqrt{P(q_* t)}},
\end{equation}
where:
\begin{equation}
P(q_* t)=(1-t^2)\left(1-\frac{\l^2 q_*^{2d}t^2}{d(d-1)}\frac{t^{2(d-1)}-1}{t^2-1}\right)
\end{equation}
We now use the expression for $q_*$ and expand everything uo to order $\l^2$. We obtain:
\begin{eqnarray}
\mathcal{P}_3(\e,d)&=&\int_{\e}^{1} h(t,d) dt+\l^2\left(a(\e,d)+\int_{\e}^{1} f(d,t) dt\right), \nonumber\\
h(t,d) &=& \frac{t^{1-d}}{\sqrt{1-t^2}}, \nonumber\\
a(\e,d)&=&\frac{\e^{2-d}}{2 d (d-1) \sqrt{1-\e^2}},\nonumber\\
f(t,d)&=&-\frac{d-2}{2 d (d-1)}\frac{t^{1-d}}{\sqrt{1-t^2}}+\frac{1}{2 d (d-1)}\frac{1-t^{2(d-1)}}{t^{d-3}(1-t^2)^{3/2}}
\end{eqnarray}
The contribution coming from $h(t,d)$ is $\l$ independent. Let's focus on the contribution coming from $f(t,d)$. We start by rewriting $f(t,d)$ as:
\begin{equation}
f(t,d)= \frac{d a(t,d)}{d t}-\frac{1}{2 d (d-1)}\frac{t^{2(d-1)}}{t^{d-3}(1-t^2)^{3/2}}.
\end{equation}
We define:
\begin{equation}
g(t,d)=\frac{1}{2 d (d-1)}\frac{t^{2(d-1)}}{t^{d-3}(1-t^2)^{3/2}}=\frac{1}{2 d (d-1)}\sum_{k=0}^{\infty}\frac{\left(3/2\right)_k}{k!} t^{2k+1+d},
\end{equation}
where we made use of a series representation for $g(t,d)$.

We want to perform the integral of $f(t,d)$. Notice that a term of $f(t,d)$ is the derivative of $a(t,d)$, however $a(t,d)$ has a singularity $t=1$, for this reason we evaluate the integral over $(\e, u)$, taking the limit $u \rightarrow 1$ in a second step.

We have:
\begin{eqnarray}
\int_{\e}^{u} f(t,d) dt= a(u,d)-a(\e,d)-\frac{1}{2 d (d-1)}\sum_{k=0}^{\infty}\frac{\left(3/2\right)_k t^{2k+2+d}}{k!(2k+1+d)}\bigg|_{\e}^{u}.
\end{eqnarray}
Notice that since $u<1$ the sum converges. By defining
\begin{equation}
\tilde a (u,d)=\frac{1}{2 d (d-1)}\sum_{k=0}^{\infty}\frac{\left(3/2\right)_k u^{2k+2+d}}{k!(2k+1+d)}
\end{equation}
we write
\begin{eqnarray}
\mathcal{P}_3(\e,d)&=&\int_{\e}^{1} h dt+\l^2\left( a(\e,d)+a(u,d)-a(\e,d)-\tilde{a}(u,d) +\frac{1}{2 d (d-1)} \sum_{k=0}^{\infty}\frac{\left(3/2\right)_k \e^{2k+2+d}}{k!(2k+1+d)} \right)\nonumber\\
&=&\int_{\e}^{1} h dt+\l^2\left(c+\frac{1}{2 d (d-1)} \sum_{k=0}^{\infty}\frac{\left(3/2\right)_k \e^{2k+2+d}}{k!(2k+1+d)}\right),\nonumber
\end{eqnarray}
where $c=\lim\limits_{u\rightarrow 1} \left( a(u,d)-\tilde{a}(u,d)\right) $. $\tilde a (t,d) $ is the primitive of $g(t,d)=\frac{1}{2 d (d-1)}\frac{t^{2(d-1)}}{t^{d-3}(1-t^2)^{3/2}}$. One finds:
\begin{equation}
\tilde a (t,d)=\frac{1}{2 d (d-1)} \frac{t^{d+2} \, _2F_1\left(\frac{3}{2},\frac{d+2}{2};\frac{d+2}{2}+1;t^2\right)}{d+2},
\end{equation}
then:
\begin{equation}
c=\frac{1}{2 d (d-1)}\frac{\sqrt{\pi } \Gamma \left(\frac{d}{2}+1\right)}{\Gamma \left(\frac{d+1}{2}\right)}.
\end{equation}
Notice that the divergent terms are $\lambda$ independent. $\l$ comes in only multiplying a constant and terms of order $\e^{2+d}$ and above (this is going to be important later).
We summarize our results by writing:
\begin{eqnarray}
\mathcal{P}_3\left(\e,d\right)&=&\mathcal{P}_0\left(\e,d\right)+\l^2(c+m_{d+2} \e^{d+2}+m_{d+4} \e^{d+4}+...)\nonumber\\
\mathcal{P}_0\left(\e,d\right)&=&\frac{n_{2-d}}{\e^{d-2}}+\frac{n_{-d}}{\e^{d}}+...+ \begin{cases} n \log \e+n_2 \e^2+...& \text{for d even}\\
n+n_1 \e+... & \text{for d odd}
\end{cases}
\end{eqnarray}
where we have made all the dependence on $\l$ explicit.
\subsubsection*{Single cut-off renormalization}
We now want to compute the following integral:
\begin{equation}
\mathcal{I}=\int_{\d/ (q_* R)}^{1}\frac{ (1-u^2)^{\frac{d-4}{2} }}{u^{d-2}}\mathcal{P}_3\left(\frac{\d}{R u},d\right) du,
\end{equation}
where we have introduced $u=Z/R$. We split this integral into three contributions we discuss separately:
\begin{eqnarray}
\mathcal{I}&=&\mathcal{I}_1+\mathcal{I}_2+\mathcal{I}_3,\nonumber\\
\mathcal{I}_1&=&\int_{\d/ (q_* R)}^{1}\frac{ (1-u^2)^{\frac{d-4}{2} }}{u^{d-2}} \mathcal{P}_0\left(\frac{\d}{R u},d\right),\nonumber\\
\mathcal{I}_2&=& \l^2 c  \int_{\d/ R}^{1}\frac{ (1-u^2)^{\frac{d-4}{2} }}{u^{d-2}} du,\nonumber\\
\mathcal{I}_3&=&  \l^2 \int_{\d/  R}^{1}\frac{ (1-u^2)^{\frac{d-4}{2} }}{u^{d-2}}\sum_{i=1}^{\infty}m_{2i+d} \left(\frac{\d}{R u}\right)^{2i+d} du \label{int2}.
\end{eqnarray}
Notice that since $\mathcal{I}_2$ and $\mathcal{I}_3$ contain an explicit factor of $\l^2$ we can take $q_*=1$.

Let's start with $\mathcal{I}_1$, by expanding $q_*^{-1}=1-\frac{\l^2}{2 d (d-1)}$ we have:
\begin{eqnarray}
\mathcal{I}_1&=&\int_{\d/ (q_* R)}^{1}\frac{ (1-u^2)^{\frac{d-4}{2} }}{u^{d-2}}\mathcal{P}_0\left(\frac{\d}{R u},d\right) du \nonumber\\
&=& \int_{\d/ R}^{1}\frac{ (1-u^2)^{\frac{d-4}{2} }}{u^{d-2}}\mathcal{P}_0\left(\frac{\d}{R u},d\right)du-\frac{\d \l^2}{2Rd(d-1)}\frac{ (1-(\d/R)^2)^{\frac{d-4}{2} }}{\d/R^{d-2}}\mathcal{P}_0\left(1,d\right) \nonumber\\
&=& \mathcal{I}_0+\l^2\sum_{i=1}N_{1-d+2i} \left(\frac{\d}{R}\right)^{-d+2i+1}.
\end{eqnarray}
Notice that $\mathcal{I}_0$ doesn't contain any $\l$ dependence, so it is going to be removed by vacuum subtraction, as we will discuss it later. Note also that in the limit $\d \rightarrow\ 0$ the logarithmic term gets killed by a $\d$ in front of it, that's why we haven't included in the sum. For $d$ odd the sum contains a constant term we don't bother to compute it since we will see later it is not universal.

We proceed now to the computation of $\mc{I}_2$. The result depends on $d$ being even or odd in particular we get:
\begin{equation}
\mc I_2=\l^2\sum_{i=1} \tilde N_{1-d+2i}\left(\frac{\d}{R}\right)^{1-d+2i}+\begin{cases}
\l^2 c\frac{\Gamma\left(\frac{3-d}{2}\right)\Gamma\left(\frac{d-2}{2}\right)}{2 \sqrt{\pi}}& \text{for $d$ even}\\ \l^2 c
\left(\frac{\left(\frac{4-d}{2}\right)_{\frac{d-3}{2}}}{\frac{d-3}{2}!}\right)\log\left(\frac{R}{\d}\right) & \text{for $d$ odd}
\end{cases}
\end{equation}
where in the case of $d$ odd we used the following expansion
\begin{equation}
(1-u^2)^{\frac{d-4}{2}}=\sum_{k=0}\frac{\left(\frac{4-d}{2}\right)_k}{k!} u^{2k},
\end{equation}
where $(a)_k$ denotes a Pochhammer symbol.

Let's now look at the last contribution, $\mc I_3$. Expanding again $(1-u^2)^{\frac{d-4}{2}}$ one can easily show that we get:
\begin{equation}
\mc I_3= \l^2 \sum_{i} M_{1-d+2i}\left(\frac{\d}{R}\right)^{1-d+2i},
\end{equation}
notice in particular integration doesn't produce any logarithmic divergence.

Summing up we have:
\begin{equation}
\mc I = \mc I_0+\l^2 \sum_{i} C_{1-d+2i}\left(\frac{\d}{R}\right)^{1-d+2i}+\begin{cases}
\l^2 c\frac{\Gamma\left(\frac{3-d}{2}\right)\Gamma\left(\frac{d-2}{2}\right)}{2 \sqrt{\pi}}& \text{for $d$ even}\\ \l^2 c
\left(\frac{\left(\frac{4-d}{2}\right)_{\frac{d-3}{2}}}{\frac{d-3}{2}!}\right)\log\left(\frac{R}{\d}\right) & \text{for $d$ odd}
\end{cases}
\end{equation}
The holographic  entanglement entropy  is obtained as:
\begin{equation}
S=\frac{L^{d-1}Vol(\mathbb{S}^{d-3})}{2 G_N} \mc I.
\end{equation}
In order to obtain physical information we need to perform a vacuum subtraction, we have:
\begin{equation}
\Delta S=\frac{L^{d-1}Vol(\mathbb{S}^{d-3})}{2 G_N} \left(\l^2 \sum_{i} C_{1-d+2i}\left(\frac{\d}{R}\right)^{1-d+2i}+\begin{cases}
\l^2 c\frac{\Gamma\left(\frac{3-d}{2}\right)\Gamma\left(\frac{d-2}{2}\right)}{2 \sqrt{\pi}}& \text{for $d$ even}\\ \l^2 c
\left(\frac{\left(\frac{4-d}{2}\right)_{\frac{d-3}{2}}}{\frac{d-3}{2}!}\right)\log\left(\frac{R}{\d}\right) & \text{for $d$ odd}\nonumber
\end{cases}\right)
\end{equation}
In particular the universal contribution is given by:
\begin{equation}\label{EEjanusd}
\Delta S_{\text{UNIV}}=\frac{L^{d-1}Vol(\mathbb{S}^{d-3})\l^2}{2 G_N} c\begin{cases}
\frac{\Gamma\left(\frac{3-d}{2}\right)\Gamma\left(\frac{d-2}{2}\right)}{2 \sqrt{\pi}}& \text{for $d$ even}\\ 
\left(\frac{\left(\frac{4-d}{2}\right)_{\frac{d-3}{2}}}{\frac{d-3}{2}!}\right)\log\left(\frac{R}{\d}\right) & \text{for $d$ odd}
\end{cases}
\end{equation}
\subsubsection*{Double cut-off renormalization}
In this renormalization procedure $\e$ is regarded as constant, we use another cut-off $Z=\d$ to regulate the integration over $Z$. We then have:
\begin{equation}
\mathcal{I}=\int_{\d/ R}^{1}\frac{ (1-u^2)^{\frac{d-4}{2} }}{u^{d-2}}\mathcal{P}\left({\e},d\right) du,
\end{equation}
with $u=Z/R$.
\begin{figure}
	\centering
	\begin{tikzpicture}
	\node at (5,3) {$\g$};
	\node at (-5,-3.5) {$\frac{\Delta S_{\text{UNIV}} 4 G_N^{(5)}}{\Vol(\mathbb{S}^1)L^3}$};
	\node at (0,0)
	{\includegraphics[width=.6\textwidth]{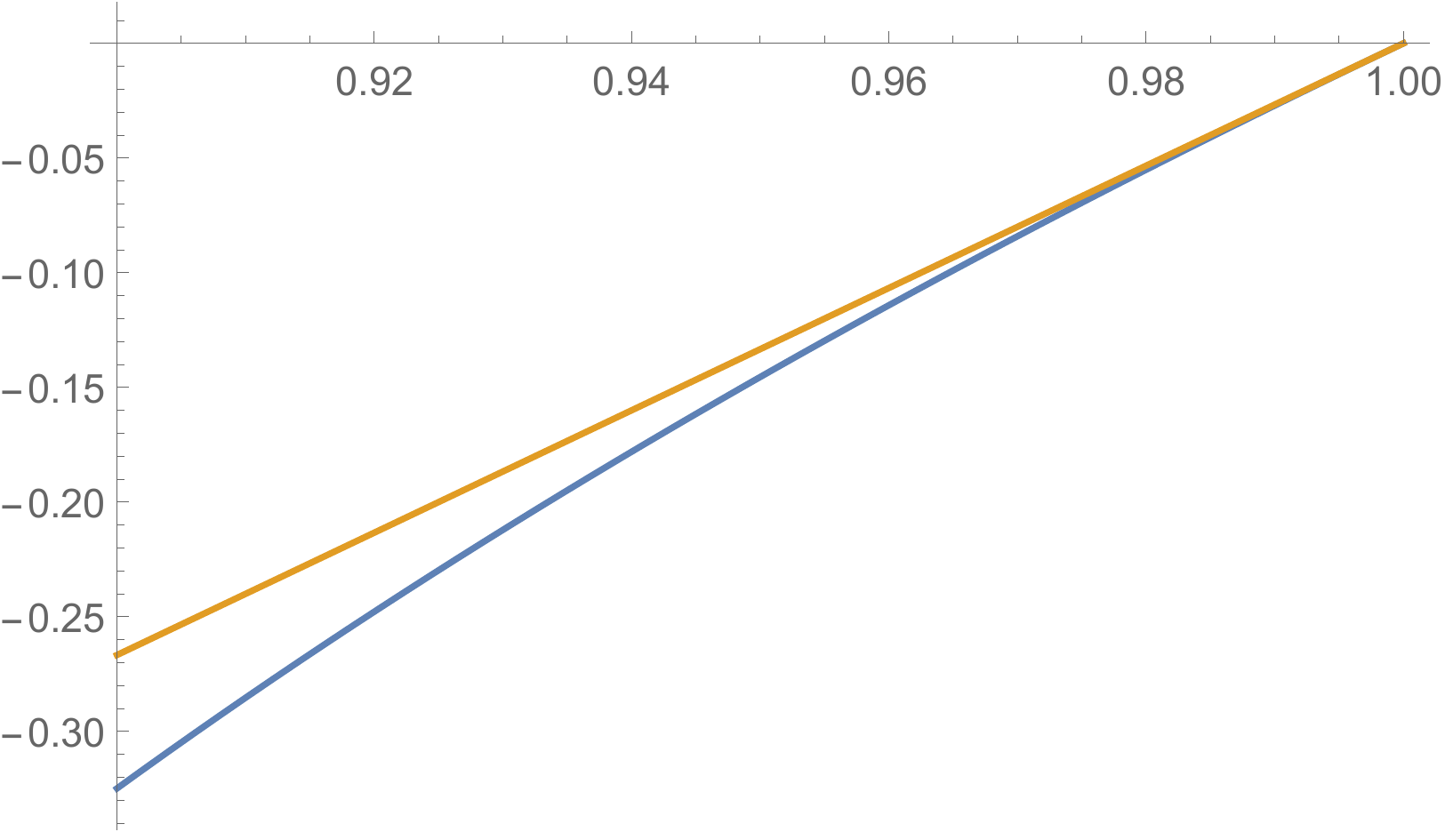}};
	\end{tikzpicture}
	\caption{The orange line represents the perturbative result for the entanglement entropy in the non Susy Janus solution computed in this appendix, while the blue line corresponds to the non perturbative computation performed in section \ref{nonsusy}. The two results agree in the perturtative regime ($\g \approx 1$).}\label{graph}
\end{figure}
Taking the difference with respect to the vacuum solution and then letting $\e \rightarrow 0$ we are left with:
\begin{equation}
\Delta \mathcal{I}=\int_{\d/ R}^{1}\frac{ (1-u^2)^{\frac{d-4}{2} }}{u^{d-2}}c \l^2 du,
\end{equation}
This is the same expression of the integral $\mc I_2$ in (\ref{int2}), which is the term containing the universal contribution. This means the two regularizations lead to the same result.\\

Notice that the $d=4$ case can be viewed as the non-supersymmetric  Janus set up studied in section \ref{nonsusy}. As a check we want to verify that taking $\gamma \rightarrow 1$ in equation (\ref{EEnonsusy}) gives (\ref{EEjanusd}) for $d=4$. Fist of all we need to find the appropriate relation between $\l$ and $\gamma$. This can be done by observing that the jump of the dilaton across the interface is a coordinate independent quantity. As we are interested in the perturbative regime we take $\l$ close to 0 and $\gamma$ close to 1, this gives\footnote{The explicit expression for the dilaton in the coordinates used in section \ref{nonsusy} can be found in \cite{D'Hoker:2007xy}. Notice that in order to compare it with equation (\ref{dilatonjanus}) we have to multiply it by a factor of $\sqrt{2}$. }:
\begin{equation}
\l^2=12 (1-\g).
\end{equation}
We can now write equation (\ref{EEjanusd}) in terms of $\gamma$:
\begin{eqnarray}
\Delta S_{\text{UNIV}}&=&\frac{\Vol(\mathbb S^1) L^3}{4 G_N^{(5)}}\left(-\frac{8}{3} (1-\g)\right)\nonumber\\
&=&\frac{\Vol(\mathbb S^1)\Vol(\mathbb{S}^5) L^8}{4 G_N^{(10)}}\left(-\frac{8}{3} (1-\g)\right),
\end{eqnarray}
where we have used dimensional reduction to relate the 5 dimensional Newton constant to the 10 dimensional one. In figure \ref{graph} we show the perturbative result derived in this section and the exact computation derived in section \ref{nonsusy}. There is agreement close as $\g$ approaches 1.

\newpage


\providecommand{\href}[2]{#2}\begingroup\raggedright\endgroup

\end{document}